\documentclass[useAMS,usenatbib,usegraphicx,onecolumn]{mn2e} 
 
\usepackage{color} 
\usepackage{psfig} 
\usepackage{amssymb} 
\usepackage{amsmath} 
\usepackage{savesym} 
\usepackage{subfigure} 
 
\newcommand{\aap}{A\&A} 
\newcommand{\mnras}{MNRAS} 
\newcommand{\pasj}{PASJ} 
\newcommand{\apjs}{ApJS} 
\newcommand{\na}{NewA} 
\newcommand{\apjl}{ApJL} 
\newcommand{\pasa}{PASA} 
\newcommand{\aj}{AJ} 
\newcommand{\apj}{ApJ}
\newcommand{\araa}{ARA\&A}
\newcommand{\nat}{Nature}
 
\newcommand{\ssa}{_a} 
\newcommand{\ssb}{_b} 
\newcommand{\ssc}{_c} 
\newcommand{\ssd}{_d} 
\newcommand{\ssi}{_i} 
\newcommand{\ssj}{_j} 
\newcommand{\sss}{_s} 
 
\newcommand{\ssab}{_{ab}} 
\newcommand{\ssac}{_{ac}} 
\newcommand{\ssad}{_{ad}}  
\newcommand{\ssai}{_{ai}}  
\newcommand{\ssaj}{_{aj}}  
\newcommand{\ssas}{_{as}}  
\newcommand{\ssat}{_{at}}  
\newcommand{\ssba}{_{ba}}  
\newcommand{\ssbc}{_{bc}}  
\newcommand{\ssbd}{_{bd}}  
\newcommand{\ssbi}{_{bi}}  
\newcommand{\ssbs}{_{bs}}  
\newcommand{\ssca}{_{ca}}  
\newcommand{\sscd}{_{cd}}  
  
\newcommand{\ssda}{_{da}}  
\newcommand{\ssia}{_{ia}}  
\newcommand{\ssij}{_{ij}}  
  
\newcommand{\ssja}{_{ja}}  
  
\newcommand{\sssa}{_{sa}}  
\newcommand{\sssb}{_{sb}}  
  
\newcommand{\sssi}{_{si}}  
  
\newcommand{\ssst}{_{st}}  
  
\newcommand{\meanhia}{\overline{h}\ssia}  
\newcommand{\meanhij}{\overline{h}\ssij}  
  
\newcommand{\meanhab}{\overline{h}\ssab}  
\newcommand{\meanhac}{\overline{h}\ssac}  
\newcommand{\meanhai}{\overline{h}\ssai}  
\newcommand{\meanhaj}{\overline{h}\ssaj}  
\newcommand{\meanhas}{\overline{h}\ssas}  
\newcommand{\meanhba}{\overline{h}\ssba}  
\newcommand{\meanhbc}{\overline{h}\ssbc}  
\newcommand{\meanhbi}{\overline{h}\ssbi}  
\newcommand{\meanhbs}{\overline{h}\ssbs}  
\newcommand{\meanhsb}{\overline{h}\sssb}  
\newcommand{\meanhsi}{\overline{h}\sssi}

\title[A hybrid SPH/N-body method for star cluster simulations]{A hybrid SPH/N-body method for star cluster simulations}  
\author[Hubber, Allison, Smith \& Goodwin]{D. A. Hubber$^{1,2}$, R. J. Allison$^{1,3}$, R. Smith$^4$, S. P. Goodwin$^1$\\  
$^{1}$Department of Physics and Astronomy, University of Sheffield, Hicks Building, Hounsfield Road, Sheffield, S3 7RH, UK \\  
$^{2}$School of Physics and Astronomy, University of Leeds, Leeds, LS2 9JT, UK \\  
$^{3}$Zentrum f\"{u}r Astronomie der Universit\"{a}t Heidelberg, Institut f\"{u}r Theoretische Astrophysik, 
Albert-Ueberle-Str. 2, 61920 Heidelberg, Germany \\  
$^{4}$Departamento de Astronomia, Universidad de Concepcion, Casilla 160-C, Concepcion, Chile}

\begin{document}  
  
\date{August 15th, 2011}  
  
\pagerange{\pageref{firstpage}--\pageref{lastpage}} \pubyear{2011}  
  
\label{firstpage}  
  
\maketitle

\begin{abstract}  
We present a new hybrid Smoothed Particle Hydrodynamics (SPH)/$N$-body  
method for modelling the collisional stellar dynamics of young 
clusters in a live gas background.  By deriving the equations of  
motion from Lagrangian mechanics we obtain a formally 
conservative combined SPH-$N$-body scheme.  
The SPH gas particles are integrated with a 2nd order Leapfrog, 
and the stars with a 4th order Hermite scheme.  
Our new approach is intended to bridge the divide between the
detailed, but expensive, full hydrodynamical simulations of star formation, 
and pure $N$-body simulations of gas-free star clusters.  We  
have implemented this hybrid approach in the SPH code SEREN  
\citep{Hubber2011} and perform a series of simple tests to demonstrate the  
fidelity of the algorithm and its conservation properties.  
We investigate and present resolution criteria to adequately resolve the density field and to prevent strong numerical scattering effects.  Future developments will include a more sophisticated treatment of binaries.
\end{abstract}  
  

\section{Introduction} \label{S:INTRO}  
The formation and dissolution of young stellar clusters is an important, but complex problem that requires computer simulations to explore in detail.  Stars form rapidly from turbulent molecular gas, most often in clusters of tens to thousands of stars \citep[e.g.][]{Elmegreen2000, LadaLada2003, McKeeOstriker2007,  Klessen2009}.  Therefore the early phases of the dynamical evolution of star clusters involves young stars moving through a significant and dynamic gaseous background.  To understand the complete process of cluster formation and evolution requires a combination of self-gravitating hydrodynamics for the gas from which stars and planets form, and the gravitational $N$-body dynamics of (multiple) stars and planets once they have formed.  

Previously, detailed hydrodynamics and accurate $N$-body dynamics have been separated.  Hydrodynamical simulations have been used to simulate the turbulent gas dynamics leading to fragmentation and star formation, while $N$-body simulations tend to follow the late gas-free stages of star cluster evolution.  However, there is a very signifcant and important phase in the life of a star cluster in which stellar dynamics within a gas background is vitally important.  In the `gas-rich' phase, which occurs around 1 -- 5~Myr\footnote{We note that star formation is not an instantaneous process and that stars are still forming whilst others are interacting and dynamically evolving.  However, we make a rough first approximation that {\em most} stars form in the first Myr, they then evolve in a dynamical gas background which is expelled at 5~Myr.} the stars are interacting dynamically in a live gas background.  The star formation process tends to produce binary and multiple systems in complex hiearchical structures.  Dynamical interactions between single and multiple systems during the subsequent few~Myr changes the binary properties of the stars as well as the structure and dynamics of the whole cluster \citep[see][and references therein]{Allison2009, Goodwin2010}.  Therefore, the binary properties of stars released into the field after gas expulsion will depend on stellar dynamics during the gas-rich phase \citep[see also][]{Kroupa1995}.  In addition, the early stages of planet formation will occur during this gas-rich phase and interactions may seriously alter the architecture and properties of planetary systems \citep[][]{Parker2011}.  Accurate observations of the binary and dynamical properties of clusters are usually only available once gas is expelled (especially those to be provided by Gaia), which means they will have been altered by dynamical evolution in the gas-rich phase.

In hydrodynamical simulations, we generally replace the dense collapse phase of gas into stars with sink particles (see \citealt{BBP1995} for SPH; \citealt{AMRsinks2004} for AMR implementations, see also \citealt{Federrath2010}).  These sink particles can represent individual stars if their sizes are $\lesssim 1$~AU \citep[e.g.][]{BBB2003, Goodwin2004, Bate2009, Offner2009} or larger regions perhaps containing primordial multiple systems which cannot be resolved if their sizes are $\gtrsim 10$~AU \citep[e.g.][]{BVB2004, SCB2009, Jappsen2005}.  Sinks have the huge advantage of allowing dense, computationally expensive, and unresolvable regions to be `compressed' into a particle which can interact with the surrounding gas and accrete from it.  However, sink particles are not point-like $N$-body particles as, even if each sink represents a single star, (a) their gravity is softened, and (b) sinks accrete from the surrounding gas.  
  
Pure $N$-body simulations of stellar systems have a long history \citep[see][]{Aarseth2003,HeggieHut}, but most ignore the early gas-rich phases of a star cluster's life.  The usual way to include gas and model the gas-rich phase is to introduce an external potential, which is often a simple Plummer or King model \citep[e.g.][]{Lada1984, Goodwin1997, BaumgardtKroupa2007, MoeckelClarke2011, Smith2011}, although see \citet{Geyer2001} used softened `star particles' in a live gas background.  Generally, the external potential is allowed to vary with time, e.g. to model the expulsion of gas from a cluster.  However, the use of a simple analytic external potential to model the gas (which is often the majority of the mass in an embedded cluster) is clearly a vast over-simplification.

In this paper we introduce a new hybrid $N$-body/Smoothed Particle Hydrodynamics (SPH) algorithm that has been implemented in the SPH code SEREN \citep{Hubber2011}.   The stellar dynamics are computed with a 4th-order integrator allowing the details of $N$-body interactions between stars to be followed.  SPH gas particles are used to represent a live background gas potential in which the stellar dynamics is modelled.
We emphasise that this hybrid method is not a replacement for fully self-consistent, high-resolution star formation simulations or detailed $N$-body simulations.  Rather, it represents a fast way of exploring stellar dynamics in a live background potential which can be used to perform large suites of simulations to explore large parameter space, 
or to inform the initial conditions of pure $N$-body simulations of the post-gas phase.  
  
In Section \ref{S:METHOD}, we introduce the hydrodynamical and N-body methods used and how they are combined algorithmically.  In Section \ref{S:TESTS}, we present a number of simple tests to demonstrate the accuracy and robustness of our method.  In Section \ref{S:DISCUSSION}, we discuss various important caveats of our method, in particular understanding resolution effects, and also discuss possible astrophysical problems that can be explored with this code.

\section{Numerical method} \label{S:METHOD}  
  
Self-gravitating hydrodynamical simulations in astrophysics are usually modelled using either a Lagrangian, particle-based approach such as Smoothed Particle Hydrodynamics \citep{Lucy1977,GM1977}, or an Eulerian, grid-based approach such as Adaptive Mesh Refinement Hydrodynamics \citep{AMR1989}.  Whereas SPH derives interaction terms by computing particle-particle force terms and integrating the motion of each particle individually, grid codes operate by computing fluxes across neighbouring grid cells.  Since N-body codes also work by computing forces and integrating positions and velocities, SPH is the most natural hydrodynamical method to merge directly with $N$-body dynamics as the particle-nature of the gas and stars are easily compatible making it straight-forward to derive the coupling force terms and to merge their individual integration schemes.  

We use a conservative self-gravitating SPH formulation \citep{PM2007} to model the gas dynamics and include the star particles within the SPH formulation as a special type of SPH particle, rather than external $N$-body particles.  Following most modern conservative SPH schemes \citep[e.g.][]{SH2002,PM2007}, the smoothing length of a gas particle $i$ is set by the relation,   
\begin{eqnarray} \label{EQN:HRHO}  
h\ssi &=& \eta\,\left( \frac{m\ssi}{\rho\ssi} \right)^{1/3}\,,  
\end{eqnarray}  
and the SPH gas density is given by   
\begin{eqnarray} \label{EQN:SPHRHO}  
\rho\ssi  &=& \sum \limits_{j=1}^{N}  m\ssj W({\bf r}\ssij,h\ssi)\,.  
\end{eqnarray}  
where ${\bf r}\ssi$, $h\ssi$, $m\ssi$, $\rho\ssi$ are the position, smoothing length, mass and density of particle $i$ respectively, ${\bf r}\ssij \equiv {\bf r}\ssi - {\bf r}\ssj$, $W$ is the SPH smoothing kernel and $\eta$ is a dimensionless number that controls the mean number of neighbours (usually set to $1.2$ to have $\sim\,60$ neighbours).  Since $h\ssi$ and $\rho\ssi$ depend on each other, we must iterate between Equations \ref{EQN:HRHO} and \ref{EQN:SPHRHO} in order to reach a consistent solution.  In contrast, the star particles have a constant smoothing length which represents the gravitational softening length to prevent violent 2-body collisions with other stars, in place of using more complicated algorithms such as regularisation \cite[See][for a description of common N-body techniques]{Aarseth2003}.  In order to reduce `scattering' during star-gas interactions, we use the mean-smoothing length approach \citep[See Appendix A of][]{PM2007}, to keep star-gas interactions as smooth as possible.  We can now formulate the Lagrangian of the system containing all interaction terms and then derive the equations of motion via the Euler-Lagrange Equations.   This simple approach allows us to develop a conservative scheme which in principle can be integrated to arbitrary accuracy (i.e. if direct summation of gravitational forces and a constant, global timestep is used).  Due to the larger energy errors often produced by N-body encounters, we use a higher-order Hermite integration scheme \citep{Hermite1992} to integrate star particles, and a simpler 2nd-order Leapfrog kick-drift-kick scheme to integrate the gas particles motion.

\subsection{Gravitational force softening in SPH} \label{SS:GRAVSOFT}  
The gravitational force softening between SPH particles can be derived in a number of ways \citep[e.g. Plummer softening, See][]{Dehnen2001}.  However, it has been suggested by \citet{BateBurkert} that it is safest to use the SPH kernel itself to derive the softening terms to prevent artificial gravitational fragmentation.  They showed that for gas condensations  where the Jeans length was of order the smoothing length or greater than, the net hydrodynamical force is stronger than the net gravitational force from all neighbouring particles, thereby suppressing or even reversing the collapse of the condensation and preventing fragmentation.  We therefore derive the softening terms from the SPH kernel following the method and nomenclature of \citet{PM2007}.  
  
First, we consider the case of uniform smoothing length.  The gravitational potential at the position of particle $i$ due to a distribution of SPH particles is given by   
\begin{eqnarray} \label{EQN:KSGRAVPOT}  
\Phi\ssa &=& G\, \sum \limits_{b=1}^{N} m\ssb \,  
\phi({\bf r}\ssab , h) \,,  
\end{eqnarray}  
where $\phi$ is the gravitational softening kernel \citep{PM2007} and $h$ is the smoothing length of all SPH particles.  We note that Eqn. \ref{EQN:KSGRAVPOT} requires the softening kernel to be a negative quantity.  The potential is related to the density field by Poisson's Equation,   
\begin{eqnarray} \label{EQN:POISSON}  
\nabla^2 \Phi({\bf r}) &=& 4\,\pi\,G\,\rho({\bf r})  
\end{eqnarray}   
where the SPH density defined at ${\bf r}\ssa$ is given by Eqn. \ref{EQN:SPHRHO}.  
This allows us to directly relate the softening kernel to the SPH smoothing kernel for a consistent formulation of self-gravity in SPH.  Substituting Equations \ref{EQN:SPHRHO} \& \ref{EQN:KSGRAVPOT} into Poisson's Equation, we obtain  
\begin{eqnarray} \label{EQN:WPHI}  
W(r,h) &=& \frac{1}{4\,\pi\,r^2} \frac{\partial}{\partial r}   
\left( r^2 \frac{\partial \phi}{\partial r}(r,h) \right)\,.  
\end{eqnarray}  
By direct integration of Eqn. \ref{EQN:WPHI} with the appropriate limits, we can obtain the gravitational softening kernel via the gravitational force kernel, $\phi'$ where  
\begin{eqnarray} \label{EQN:GRAVFORCEKERNEL}  
\phi'(r,h) &\equiv& \frac{\partial \phi}{\partial r}(r,h) = \frac{4\,\pi}{r^2}\int\limits_0^r{W(r',h)\,r'^2\,dr'}\,,\label{EQN:GRAVPOTKERNEL}  
\end{eqnarray}  
and   
\begin{eqnarray}  
\phi(r,h) &=& 4\,\pi\left(-\frac{1}{r}\int\limits_0^r{W(r',h)\,r'^2\,dr'}+\int\limits_0^r{W(r',h)\,r'\,dr'}-\int\limits_0^{{\cal R}h}{W(r',h)\,r'\,dr'}\right)\,.  
\end{eqnarray}
where ${\cal R}$ is the compact support of the kernel (e.g. ${\cal R} = 2$ for the M4-kernel).
  
When using variable smoothing lengths, we have two choices for symmetrising the gravitational interaction; (a) use the average of the two softening kernels, or (b) use the mean smoothing length in the softening kernel, i.e.   
\begin{eqnarray}  
\Phi\ssa = G\, \sum \limits_{b=1}^{N} m\ssb \,  
\frac{\phi({\bf r}\ssab , h\ssa) + \phi({\bf r}\ssab , h\ssb)}{2}  \,, & \;\;\;\;{\rm or}\;\;\;\; &  
\Phi\ssa = G\, \sum \limits_{b=1}^{N} m\ssb \,  
\phi({\bf r}\ssab , \meanhab)\,,  
\end{eqnarray}  
where $\meanhab \equiv \frac{1}{2}\,(h\ssa + h\ssb)$.  \citet{PM2007} advocate using the average softening kernel approach since it requires less loops over all particles than the mean smoothing length approach.  When using only SPH particles where the smoothing length is determined by Eqn. \ref{EQN:HRHO}, there is little difference in the results since the two methods give similar potentials and forces.  However, when we include star particles which can have an arbitrary small smoothing length, there can be very large discrepancies between the two methods.  For example, consider a `collision' between a gas particle and an SPH particle, i.e. where the two particle lie at almost the same position in space.  The mean softening kernel approach has two terms, one which will be quite small due to the smoothing length of the gas particle, and the second term using the star smoothing length which can become very large producing a corresponding large scattering force.  The mean smoothing length approach however, can never produce a large scattering force, even if the smoothing length of the star becomes zero since the average of the two smoothing lengths can never be less than $\frac{1}{2}\,h\ssi$ or $\frac{1}{2}\,h\ssj$.  Furthermore, the mean smoothing length method allows us to use star particles with zero smoothing length, permitting the study of truly collisional stellar dynamics with unsoftened star-star forces, but softened star-gas interactions.  Therefore, for the case of our hybrid formulation, we advocate using the mean smoothing length approach and we derive all subsequent equations of motion using this method.

\subsection{Conservative SPH} \label{SS:GRADHSPH}  
The SPH equations of motion can be derived from Lagrangian mechanics, resulting in a set of equations that automatically conserve linear momentum, to machine precision, and angular momentum and energy, both to integration error \citep[See][]{SH2002,PM2007}.   \citet{PM2007} derived the equations of motion for self-gravitating systems with variable smoothing lengths using Lagrangian mechanics.  Following their method, we derive the equations of motion for a set of SPH particles (with variable smoothing length given by Eqns. \ref{EQN:HRHO} \& \ref{EQN:SPHRHO}) and stars (with a fixed smoothing length).    
If we have $N_g$ gas particles with labels $b = 1,2, ... ,N_g$ and $N_s$ star particles with labels $i = 1,2, ... ,N_s$, then by inserting all terms into the Lagrangian, we obtain   
\begin{eqnarray} \label{EQN:LAGRANGIAN}  
{\cal L} &=& \frac{1}{2}\sum \limits_{b=1}^{N_g} {m\ssb \, v\ssb^2 }  + \frac{1}{2}\sum \limits_{i=1}^{N_i} {m\ssi\,v\ssi^2 } - \sum \limits_{b=1}^{N_g} {m\ssb\,u\ssb} + {\cal L}_{_{\rm GRAV}}  
\end{eqnarray}  
where ${\cal L}_{_{\rm GRAV}}$ is the gravitational contribution to the Lagrangian given by   
\begin{eqnarray} \label{EQN:GRAVLAGRANGIAN}  
{\cal L}_{_{\rm GRAV}} &=&   
- \frac{G}{2} \sum \limits_{b=1}^{N_g} \, \sum \limits_{c=1}^{N_g}\, { m\ssb m\ssc \phi\ssbc(\overline{h}\ssbc) } -   
\frac{G}{2} \sum \limits_{i=1}^{N_s} \, \sum \limits_{j=1}^{N_s}\, { m\ssi m\ssj \phi\ssij(\overline{h}\ssij) }   
- G \sum \limits_{b=1}^{N_g} \, \sum \limits_{i=1}^{N_s}\, { m\ssb m\ssi \phi\ssbi(\overline{h}\ssbi) } \,.  
\end{eqnarray}  
We note that throughout this paper, summations over all SPH gas particles are given by the indices $b$, $c$ and $d$ and summations over all star particles by $i$, $j$ and $k$.  The equations of motion can be obtained by inserting this Lagrangian into the Euler-Lagrange equations,   
\begin{eqnarray} \label{EQN:EULERLAGRANGE}  
\frac{d}{dt} \left( \frac{\partial {\cal L}}{\partial {\bf v}\ssa} \right)   
- \frac{\partial {\cal L}}{\partial {\bf r}\ssa} &=& 0 \,.  
\end{eqnarray}  
After inserting the correct terms and evaluating the algebra (See Appendix \ref{A:SPHDERIVATION} for a full derivation),  we obtain the following equation of motion for SPH gas particles   
\begin{eqnarray} \label{EQN:SPHEOM}  
{\bf a}\ssa &=&   
- \sum \limits_{b=1}^{N_g}   
m\ssb \,\left[ \frac{P\ssa}{\rho^2\ssa\Omega\ssa}\frac{\partial W\ssab}{\partial {\bf r}\ssa}(h\ssa)  + \frac{P\ssb}{\rho^2\ssb\Omega\ssb}\frac{\partial W\ssab}{\partial {\bf r}\ssa}(h\ssb) \right]  \nonumber \\
&& - G \sum \limits_{b=1}^{N_g} m\ssb\, \phi'\ssab(\overline{h}\ssab)\,\hat{\bf r}\ssab   
- G \sum \limits_{i=1}^{N_s} \,m\ssi\,\phi'\ssai(\overline{h}\ssai)\,\hat{\bf r}\ssai   
- \frac{G}{2} \sum \limits_{b=1}^{N_g}   
m\ssb \,\left[ \frac{\left(\bar{\zeta}\ssa + \bar{\chi}\ssa \right)}{\Omega\ssa}\frac{\partial W\ssab}{\partial {\bf r}\ssa}(h\ssa)  + \frac{\left(\bar{\zeta}\ssb + \bar{\chi}\ssb \right)}{\Omega\ssb}\frac{\partial W\ssab}{\partial {\bf r}\ssa}(h\ssb) \right]  
\end{eqnarray}  
where $P\ssa = (\gamma - 1)\,\rho\ssa\,u\ssa$ is the thermal pressure of particle $a$, and $\Omega\ssa$, $\bar{\zeta}\ssa$, and $\bar{\chi}\ssa$ are defined by   
\begin{eqnarray} \label{EQN:OMEGA}  
\Omega\ssa = 1 - \frac{\partial h\ssa }{\partial \rho\ssa }  
\sum \limits_{b=1}^{N}  m\ssb  \frac{\partial W\ssab}{\partial h}  
(h\ssa )\,.  
\end{eqnarray}  
\begin{eqnarray}  
\bar{\zeta}\ssa = \frac{\partial h\ssa}{\partial \rho\ssa}   
\sum \limits_{b=1}^{N} m\ssb \frac{\partial \phi\ssab}{\partial \meanhab}(\overline{h}\ssab)\,,  
\label{EQN:GRADHZETA}  
\end{eqnarray}  
\begin{eqnarray}  
\bar{\chi}\ssa = \frac{\partial h\ssa}{\partial \rho\ssa}   
\sum \limits_{i=1}^{N} m\ssi \frac{\partial \phi\ssai}{\partial \meanhai}(\overline{h}\ssai)\,.
\label{EQN:GRADHCHI}  
\end{eqnarray}  
The $\Omega$ term is the familiar `grad-h' correction term that appears in conservative SPH with varying smoothing length \cite[e.g.][]{SH2002,PM2007}. The $\bar{\zeta}$ term is the correction term derived by \citet{PM2007} for gravitational interactions between gas particles in conservative SPH.  We obtain an analogous correction term for the star-gas interaction, $\bar{\chi}$, which is a summation over all neighbouring star particles.  However, $\bar{\chi}$ is still included in a summation over all neighbouring gas particles since it is the variation in the smoothing length (which is determined by neighbouring particle positions) that gives rise to the correction terms.  We have some choice in how to evolve the thermal properties of the gas particles.  We chose to evolve the specific internal energy equation, i.e. 
\begin{eqnarray}
\frac{du\ssa}{dt} = \frac{P\ssa}{\rho^2\ssa\,\Omega\ssa}\,
\sum \limits_{b=1}^{N_g} { m\ssb\,{\bf v}\ssab \cdot \frac{\partial W\ssab}{\partial {\bf r}\ssa} }
\end{eqnarray}

For the star particles, we obtain the following equations of motion for star $s$,
\begin{eqnarray} \label{EQN:STARGRAV}  
{\bf a}\sss &=& -G \sum \limits_{i=1}^{N_s} {  
m\ssi\,\phi'\ssst(\overline{h}\sssi)\,\hat{\bf r}\sssi} \,   
- G \sum \limits_{b=1}^{N_g}\,{m\ssb\,\phi'\sssb(\overline{h}\sssb)\,\hat{\bf r}\sssb } \,.  
\end{eqnarray}  
We note that since only the smoothing lengths of the gas particles are allowed to vary, all correction terms derived via the Lagrangian appear in the equations of motion for the gas particles.

\subsection{Coupled-integration scheme}  
The equations of motion for both gas and star particles can be integrated either with a single integration scheme, or with two independent integration schemes in parallel.  Current SPH codes typically use 2nd-order schemes, such as the Leapfrog or the Runge-Kutta-Fehlberg, whereas N-body codes use at least 4th-order schemes such as the Hermite scheme.  In our implementation, we chose to use a 2nd-order Leapfrog kick-drift-kick scheme \citep[][]{Gadget2} to integrate the SPH gas particles coupled with a 4th-order Hermite integration scheme \citep[][]{Hermite1992} to integrate the star particles.  One important reason for this choice is that a 4th-order Hermite scheme can be considered as the higher-order equivalent of the leapfrog scheme \citep[See][]{Hut1995}, where the force, prediction and correction steps are all computed at the same points in the timestep for both schemes.  We discuss the details of our implementation of both integration schemes, and in particular we discuss the modifications to the 4th-order Hermite scheme to include SPH smoothing.

\subsubsection{2nd-order SPH leapfrog integration scheme} \label{SSS:LEAPFROG}  
We integrate the SPH particles using a 2nd-order Leapfrog kick-drift-kick integration scheme.  A traditional leapfrog works by advancing the positions and velocities half-a-step apart, i.e.   
\begin{eqnarray} \label{EQN:LEAPFROG}  
{\bf v}\ssa^{n+1/2} &=& {\bf v}\ssa^{n-1/2} + {\bf a}\ssa^{n}\,\Delta t \,,\\  
{\bf r}\ssa^{n+1} &=& {\bf r}\ssa^{n} + {\bf v}\ssa^{n+1/2} \,\Delta t\,.  
\end{eqnarray}
It is possible to transform the traditional leapfrog equations into a form where the positions and velocities are both updated at the end of the step,   
\begin{eqnarray}  
{\bf r}\ssa^{n+1} &=& {\bf r}\ssa^{n} + {\bf v}\ssa^{n}\,\Delta\,t + \frac{1}{2}\,{\bf a}\ssa^{n}\,\Delta\,t^2\,, \label{EQN:RLF2} \\  
{\bf v}\ssa^{n+1} &=& {\bf v}\ssa^{n} + \frac{1}{2}\,\left( {\bf a}\ssa^{n} + {\bf a}\ssa^{n+1} \right)\,\Delta t \,. \label{EQN:VLF2}  
\end{eqnarray}  
This form of the leapfrog (also known as the Velocity-Verlet integration scheme) has 3rd-order accuracy in integrating the positions and 2nd-order accuracy in integrating the velocities, and therefore 2nd-order overall.  As we will see in Section \ref{SSS:HERMITE}, it also has the useful property of having its acceleration and `correction' terms calculated in sync with the corresponding terms for the 4th order Hermite scheme used for integrating the star particles.  Other integration schemes (e.g. the 2nd-order Leapfrog drift-kick-drift) do not necessarily share this property.  
  
The timesteps are determined by taking the minimum of three separate conditions, the SPH Courant-Friedrichs-Lewy condition \citep{CFL1928}, the acceleration condition, and the energy condition when cooling is employed \citep[See][]{Hubber2011}, i.e.  
\begin{eqnarray}
\Delta t\ssa &=& {\rm MIN}\,\, \left\{   
\frac{\gamma_{_{\rm COUR}}\,h\ssa}{(1 + 1.2\,\alpha)\,c\ssa + (1 + 1.2\,\beta)\,h\ssa\,
|\nabla \cdot {\bf v}\ssa|}\,,\,  
\left(\frac{\gamma_{_{\rm ACCEL}}\,h\ssa}{|{\bf a}\ssa| + \epsilon} \right)^{1/2} \,,\,  
\frac{\gamma_{_{\rm ENERGY}}\,u\ssa}{|\dot{u} \cdot \ssa| + \epsilon} \, \right\}  
\end{eqnarray}  
where the $\gamma$ terms are dimensionless timestep multipliers and $\epsilon \ll 1$ is a small number used to prevent a divide-by-zero in the case of $|{\bf a}\ssa| = 0$ or $|\dot{u}\ssa| = 0$.

\begin{figure}  
\centerline{\psfig{figure=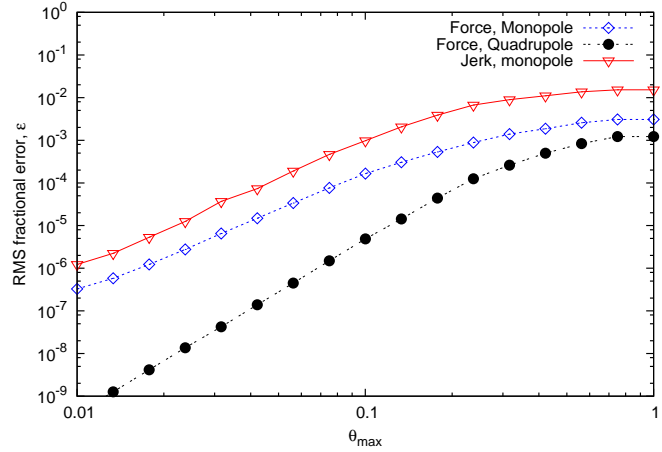,width=9.0cm,angle=270}}  
\caption{The RMS fractional error using tree gravity of (a) gravitional force with monopole terms only (blue diamonds), (b) gravitational force including the quadrupole terms (solid black circles), and (c) gravitational `jerk' with monopole terms only (red triangles) using the geometric MAC as a function of opening angle, $\theta_{_{\rm MAX}}$.}
\label{FIG:JERKERROR}
\end{figure}

\subsubsection{4th-order N-body Hermite integration scheme} \label{SSS:HERMITE}  
We use a standard 4th-order Hermite integrator \citep{Hermite1992} modified by including the same SPH softening scheme as the gas particles \citep{Hubber2011} to integrate the motion of the star particles.  
In the Hermite scheme, the first time derivative of the acceleration (often referred to as the `jerk') must be calculated explicitly from Equation \ref{EQN:STARGRAV}.  
By taking the time derivative of Equation \ref{EQN:STARGRAV} and using Equation \ref{EQN:WPHI}, we obtain for the jerk term,
\begin{eqnarray}  
\dot{\bf a}^n\sss &=& -\,G\,\sum \limits_{i=1}^{N}   
{\frac{m\ssi\,\phi'({\bf r}\sssi , \meanhsi )}{|{\bf r}\sssi |}{\bf v}\sssi} \; + \;3\,G\, \sum \limits_{i=1}^{N}  
{\frac{m\ssi\,({\bf r}\sssi  \cdot {\bf v}\sssi)\,\phi'({\bf r}\sssi , \meanhsi )}{|{\bf r}\sssi |^3} {\bf r}\sssi } \;   
- \;4\,\pi\,G\, \sum \limits_{i=1}^{N} {\frac{m\ssi\,\,({\bf r}\sssi  \cdot {\bf v}\sssi)\,   
W({\bf r}\sssi , \meanhsi )}{|{\bf r}\sssi |^2}{\bf r}\sssi}\, \nonumber \\
&& -\,G\,\sum \limits_{b=1}^{N}   {\frac{m\ssb\,\phi'({\bf r}\sssb , \meanhsb )}{|{\bf r}\sssb |}{\bf v}\sssb} \; + \;3\,G\, \sum \limits_{b=1}^{N}  
{\frac{m\ssb\,({\bf r}\sssb  \cdot {\bf v}\sssb)\,\phi'({\bf r}\sssb , \meanhsb )}{|{\bf r}\sssb |^3} {\bf r}\sssb } \;   
- \;4\,\pi\,G\, \sum \limits_{b=1}^{N} {\frac{m\ssb\,\,({\bf r}\sssb  \cdot {\bf v}\sssb)\,   
W({\bf r}\sssi , \meanhsb )}{|{\bf r}\sssb |^2}{\bf r}\sssb}\,.  
\label{EQN:JERK}  
\end{eqnarray}  
Since the jerk can be computed from a single sum over all particles, we can compute it explicitly at the same time as computing the accelerations.  Once both terms are computed, we can calculate the predicted positions and velocities of the stars at the end of the steps, i.e.   
\begin{eqnarray}  
{\bf r}\sss^{n+1} &=& {\bf r}\sss^{n} + {\bf v}\sss^{n}\,\Delta t + \frac{1}{2}{\bf a}\sss^{n}\,\Delta t^2 + \frac{1}{6}\dot{\bf a}\sss^{n}\,\Delta t^3 \,, \\  
{\bf v}\sss^{n+1} &=& {\bf v}\sss^{n} + {\bf a}\sss^{n}\,\Delta t + \frac{1}{2}\dot{\bf a}\sss^{n}\,\Delta t^2\,.  
\end{eqnarray}  
The acceleration and jerk are then recomputed at the end of the step, ${\bf a}^{n+1}$ and $\dot{\bf a}^{n+1}$.  This then allows us to compute the higher order time derivatives,  
\begin{eqnarray}  
\ddot{\bf a}\sss^{n} &=& \frac{2 \left( -3({\bf a}\sss^{n} - {\bf a}\sss^{n+1}) - (2\dot{\bf a}\sss^{n} + \dot{\bf a}\sss^{n+1})\Delta t \right)}{\Delta t^2}\,,  \label{EQN:A2} \\  
\dddot{\bf a}\sss^{n} &=& \frac{6 \left( 2({\bf a}\sss^{n} - {\bf a}\sss^{n+1}) + (\dot{\bf a}\sss^{n} + \dot{\bf a}\sss^{n+1})\Delta t \right)}{\Delta t^3}\,. \label{EQN:A3}  
\end{eqnarray}  
Finally, we apply the correction step where the higher-order terms are added to the position and velocity vectors, i.e.   
\begin{eqnarray}  
{\bf r}\sss^{n+1} &=& {\bf r}\sss^{n+1} + \frac{1}{24}\ddot{\bf a}\sss^{n}\,\Delta t^4 + \frac{1}{120}\,\dddot{\bf a}\sss^{n}\,\Delta t^5 \, \\  
{\bf v}\sss^{n+1} &=& {\bf v}\sss^{n+1} + \frac{1}{6}\ddot{\bf a}\sss^{n}\,\Delta t^3 + \frac{1}{24}\,\dddot{\bf a}\sss^{n}\,\Delta t^4 \,.  
\end{eqnarray}  
  
We compute the N-body timesteps using the Aarseth timestep criterion \citep{Aarseth2003},   
\begin{eqnarray}  
\Delta t\sss &=& \gamma\sss \, \sqrt{\frac{|{\bf a}\sss| |\ddot{\bf a}\sss| +   
|\dot{\bf a}\sss|^2}{|\dot{\bf a}\sss| |\dddot{\bf a}\sss| +   
|\ddot{\bf a}\sss|^2}}\,. \label{EQN:DTAARSETH}  
\end{eqnarray}  
where $\gamma\sss$ is the timestep multiplier for stars.  For the very first timestep, we must compute the 2nd and 3rd time derivatives explicitly \citep{Aarseth2003} since we do not yet have information on the 2nd and 3rd derivatives (since they are only first computed at the end of the first timestep).  Hereafter, we use Eqns. \ref{EQN:A2} \& \ref{EQN:A3} for computing these derivatives.

\subsection{Calculating gravitational terms}
SEREN \citep{Hubber2011} uses a Barnes-Hut gravity tree \citep{BHtree} for computing gravitational forces for all self-gravitating gas particles.  We use the same tree for computing the gravitational forces due to all gas particles for both SPH gas particles, and star particles.  SEREN has a variety of tree-opening criteria that can be selected at compilation-time.  For most simulations in this paper, we use the Eigenvalue multipole-acceptance criterion \citep{Hubber2011} because it has better force error control, and therefore ultimately better energy error control, than the standard geometric opening-angle criterion often used.  For nearby SPH particles, which require direct computation of the gravitational acceleration, we also compute the jerk term when computing the acceleration of star particles.  For tree cells, we compute the jerk contribution due to the centre of mass of the cell; however, we ignore the contribution due to the quadrupole moment terms for simplicity.  For forces due to star particles, we sum all the contributions for the gravitational acceleration (and jerk) directly without using a gravity tree.  

In Figure \ref{FIG:JERKERROR}, we plot the RMS force (monopole and quadrupole) and jerk (monopole only) errors using the geometric opening-angle criterion (in order to plot both monopole and quadrupole errors) for stars in a star-gas Plummer sphere (as discussed in Section \ref{SS:PLUMMERTEST}).  We can see that the jerk error scales in the same way as the monopole-only force error.  The force error using quadrupole moments scales much better than both monopole errors, as expected.  In order to allow high-accuracy in the calculation of the jerk while still using the tree, we use two different tree opening criteria; one for SPH gas particles (which do not need to calculate the jerk), and a stricter one for N-body particles that use the tree.  It should be noted that Figure \ref{FIG:JERKERROR} represents an upper limit to the expected jerk error.  The dominant contribution to the jerk will be from close encounters with other stars, which is computed exactly.

\subsection{Block timestepping}
SEREN \citep{Hubber2011} uses a standard block-timestepping scheme used in many N-body and SPH codes.  The timesteps of all gas and star particles are restricted to being $\Delta t = \Delta t_{_{\rm MAX}}\,/\,2^{n}$ where $n = 0,1,2,3,4,..$ is a positive integer.  All particles and stars therefore are all exactly synchronised on the longest timestep, when the timestep level structure is recomputed.  In the standard block-timestepping scheme, particle timesteps are only computed once their current timestep has been completed.  At the end of the step, particles are allowed to move to any lower timestep (higher $n$), or are allowed to move up one level (lower $n$) provided the new higher level is synchronised with the old level.  This approach means particles can rapidly and immediately drop to short timesteps when required, but are only allowed to rise back up slowly to prevent timesteps oscillating up and down too frequently. 
 
SEREN also contains the neighbour-timestep monitoring procedure of \citet{SM2009} to prevent large timestep differences between SPH neighbours generating large energy, momentum and angular momentum errors.  Although this algorithm is not necessarily needed for the tests presented in this paper, it will almost certainly be required for future applications where feedback processes can suddenly generate large discontinuities in density and temperature, resulting in large timestep disparities.

\section{Tests} \label{S:TESTS}  
  
We present a small suite of numerical tests demonstrating the accuracy of our hybrid SPH/$N$-body approach.  Tests using the SPH and $N$-body components independently were presented by \citet{Hubber2011}.  The SPH component was tested using a variety of shock-tube, Kelvin-Helmholtz instability and gravity tests, as well as tests of the tree and the error-scaling of the code.  The $N$-body component was tested with some 3-body examples that had known solutions.  In this paper, we present tests of the combined scheme to demonstrate that the conservative equations of motion derived in Section \ref{SS:GRADHSPH} are correct, and that the scheme does not exhibit any unexpected numerical effects.  We perform a simple test investigating the scattering between gas and star particles.  Using star-gas Plummer spheres, we test the expected error scalings of the SPH and $N$-body components and the combined scheme, as well as the stability of the star-gas Plummer spheres.  Finally, we perform a simple test of colliding star-gas Plummer spheres.  In all tests we use an adiabatic equation of state in which heating and cooling are only due to $P$d$V$ work by expansion or contraction, thus enabling us to test the energy conservation of the code.  We work in dimensionless units throughout, where $G = 1$.

\subsection{Star-gas particle scattering} \label{SS:SCATTERINGTEST}  

Our hybrid SPH/$N$-body method enables investigation of stellar dynamics within a gas potential that may be time-evolving and irregular.  For this goal, we must clearly understand the origin and impact of any numerical effects on the results of our simulations.  One important difference between gas-only interactions and star-gas interactions in this scheme is that star particles are allowed to `pass through' SPH gas particles, whereas artificial viscosity will prevent gas particle penetration by forming a shock.  Even though the gravitational interactions between stars and gas are smoothed, the stars are still interacting with a gravitational field defined by discrete points and thus can be deflected by those points.  Therefore, gas particles can in principle scatter star particles significantly if the resolution is too coarse.  This test is designed to investigate how significantly gas particles can scatter star particles and to help determine resolution criteria to prevent significant numerical scattering.

For point particles obeying Newton's gravitational law, the scattering angle of a star of mass $m_s$ due to a hyperbolic encounter with a gas particle of mass $m_g$ in the centre-of-mass frame is given by  
\begin{eqnarray} \label{EQN:SCATTERANGLE} 
\Delta\,\theta_{\rm D} &=& \frac{2\,m_g}{m_g + m_s}\,\tan^{-1}{ \left( \frac{G\,(m_g + m_s)}{b\,v^2} \right) }  
\approx \frac{2\,G\,m_g}{b\,v^2} 
\end{eqnarray} 
\citep[e.g.][]{GalacticDynamics2} where $b$ is the impact parameter, $v$ is the relative tangential velocity at infinity, and the approximation is for small deflections.  For smoothed gravity, we would expect that the net deflection angle would be reduced by smoothing, and for the dependence on $b$ to be fundamentally altered since the gravitational force for neighbouring particles tends to zero as the distance becomes zero.  For the M4-kernel, the gravitational force reaches a maximum at around $|\Delta\,{\bf r}| \sim 0.8\,h$ \citep[see Figure 1 of][]{PM2007} and then decreases to zero (instead of $\propto 1/r^2$).  Therefore, an approximation to the maximum possible scattering angle can be obtained by setting $b \sim h$ in Equation \ref{EQN:SCATTERANGLE}\footnote{A more rigorous derivation involving the form of the force kernel would reveal the exact dependence of the scattering angle on $b$ and $v$.  However, setting $b = C\,h$ gives the same qualitative behaviour averaged over all impact parameters where $C$ is some multiplicative constant of order unity.}.  Since the smoothing length is dependent solely on the gas particle positions, the strength of the star-gas interaction becomes solely a function of relative velocity, $v$.  From Equation \ref{EQN:SCATTERANGLE}, we can define a critical relative velocity, $v_{\rm crit}$, where the interaction results in a significant deflection angle, where  
\begin{eqnarray} \label{EQN:VCRIT}
v_{\rm{crit}} &=& \left(\frac{2\,G\,m_g}{h} \right)^{1/2}\,.  
\label{EQN:VCRIT}  
\end{eqnarray}  
\noindent 
In this form, the deflection angle is simply $\Delta \theta_{\rm D} = v^2_{\rm crit} / v^2$.  Therefore, a simple resolution condition that ensures star-gas scattering is negligible, i.e. $\Delta \theta_{\rm D} \ll 1$, is given by $v^2 \gg v^2_{\rm crit}$.  This application of this criterion to various astrophysical scenarios will be discussed later in the paper.

In order to test the validity of this assertion, we perform simulations of a single star that is moving with velocity $v$ through a periodic gas cube of side-length $L$ and uniform density $\rho$ where the gas velocity is fixed to zero everywhere.  If the gas density field is perfectly uniform (i.e. in the continuum limit where $N_g \rightarrow \infty$), then the net gravitational force due to the gas will be zero and the star will simply move with constant velocity without any deviation or deceleration.  If the density field is not perfectly uniform, as is the case when represented by a discrete set of particles, then small deflections will alter the path of the star.  In our test, the gas particles are first relaxed to a glass \citep[See][for details]{Hubber2011} which is the most uniform arrangement of particles used in typical simulations.  We use periodic boundary conditions combined with Ewald gravity \citep{Ewald1991} to produce a net zero gravitational field in the gas, with the exception of the small deviations due to the smoothed particle distribution.   \citet{superbox2000} performed a similar scattering test to quantify the effects of numerical relaxation, although in the context of large-scale galaxy simulations.

\begin{figure}  
\centerline{\psfig{figure=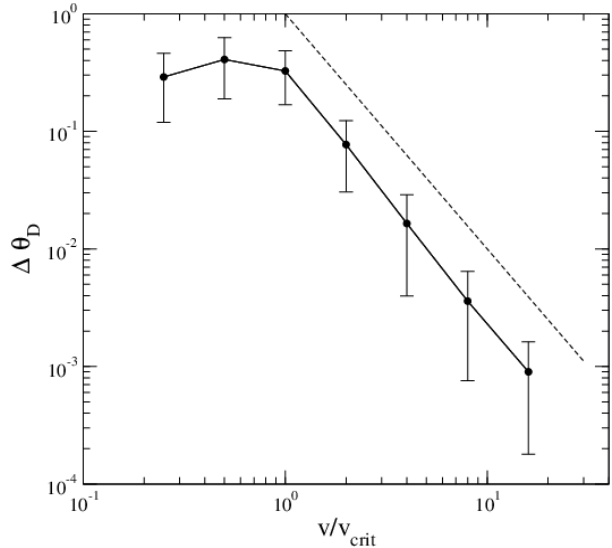,width=8.0cm,angle=0}}  
\caption{Deflection angle $\Delta \theta_{\rm{D}}$ versus star particle velocity (normalised by the critical velocity). As v increases above $v_{\rm{crit}}$, the star particle is increasingly less scattered. The scattering angle is roughly proportional to $v^{-2}$. However, for $v \le v_{\rm{crit}}$, there is a clear change in behaviour. Scattering is strong, and the scattering angle no longer follows the same power-law trend with particle velocity.}  
\label{FIG:GLASSSCATTER}  
\end{figure}

The initial velocity $v$ of the star particle is varied between simulations. We test $v=\frac{1}{4}\,v_{\rm{crit}}$, $\frac{1}{2}\,v_{\rm{crit}}$, $v_{\rm{crit}}$, $2\,v_{\rm{crit}}$, $4\,v_{\rm{crit}}$, $8\,v_{\rm{crit}}$, and $16\,v_{\rm{crit}}$. For each case, we place the particle at a random position in the gas cube in order to remove the systematic effects of using the same glass arrangement.   We simulate 6 realisations for each of the selected velocities and take the average and standard deviation of the results.  Each star particle moves along its trajectory until $t = L/v$, i.e. the crossing time of the cube.  This ensures, in so far as is possible, that during a simulation a star encounters approximately equal number of gas particles for all choices of initial velocity.  To quantify the effect of the scattering on the star particle, we measure the deflection angle, $\Delta \theta_{\rm{D}}$, i.e. the angle of the particle's trajectory relative to the initial velocity along the x-axis.  The results of these tests can be seen in Fig. \ref{FIG:GLASSSCATTER}.  
  
For $v > v_{\rm{crit}}$, increasing the star particle velocity decreases the net effect of scattering due to the gas particles.  In this regime, the scattering angle, $\Delta \theta_{\rm{D}}$, falls as $v^{-2}$, the same dependency as suggested by Eqn. \ref{EQN:SCATTERANGLE} with $b \sim h$.  We notice that the net average scattering angle is lower than expected (Figure \ref{FIG:GLASSSCATTER}; dashed line) due to the effects of smoothing.  We note the net deflection is the accumulation of several ($\sim 10$) deflections, not necessarily in the same direction hence it will `random-walk' with each deflection.  The error bars are due to the combination of the random-walk errors and the impact parameter dependence (which is reduced, but not eliminated).
  
However, for $v \leq v_{\rm{crit}}$, extremely strong scattering occurs and dominates the dynamics of the star.  At these velocities, the power-law relation between the scattering angle 
and velocity is broken since the interaction is now effectively a parabolic or elliptical interaction instead of a hyperbolic encounter.  The final deflection angle is almost random due to the strong nature of the perturbations.  The 
kinetic energy of the star is smaller than the gravitational potential energy, even accounting for smoothing, and therefore a star can in 
principle become trapped and bound to individual SPH particles.  
We note that this would not necessarily happen in a more realistic astronomical simulation because the gas particles in this test are static (so the star moves as a test particle), and therefore cannot be scattered off the star themselves.  At such velocities, the coarseness of the SPH particle distribution 
clearly introduces potentially serious numerical effects which could corrupt 
any hybrid simulation.  In future sections in this paper, we discuss possible gas resolution criteria as a means of avoiding unphysical scattering effects.

\subsection{Stability of star-gas Plummer spheres} \label{SS:PLUMMERTEST}  
  
The Plummer sphere \citep{Plummer1911} is commonly used in stellar dynamics simulations as it is described by simple, analytic formula.  For a Plummer sphere of mass, $M$, and characteristic Plummer radius, $a$, the density distribution, $\rho(r)$, is given by  
\begin{eqnarray} \label{EQN:PLUMMERRHO}  
\rho(r) &=& \frac{3M}{4 \pi a^3} \,  
\left( 1 + \frac{r^2}{a^2} \right)^{-5/2}\,,  
\end{eqnarray}  
and the 1-D velocity dispersion, $\sigma(r)$, is given by  
\begin{eqnarray} \label{EQN:PLUMMERSIGMA}  
\sigma^2(r) &=& \frac{GM}{6\,a}\left(1 + \frac{r^2}{a^2}\right)^{-1/2}\,.  
\end{eqnarray}  
We simulate Plummer spheres that are purely stars, purely gas, or a  
mixture of stars and  
gas\footnote{Note that Plummer spheres are formally infinite in  
  extent.  In practice we truncate our Plummer spheres at a radius of  
  $20\,a$.  This means that they are not in exact equilibrium; however  
  this has a negligible effect on their evolution}.  
For setting up the stellar component of our clusters, we use the method outlined by \cite{Aarseth1974}.

For the gas distributions, the Plummer model corresponds to a $n = 5$  
polytrope.  A polytrope is a self-gravitating gas whose equation of  
state obeys the form $P = K\,\rho^{1 + \frac{1}{n}}$ and whose density  
structure is a solution of the Lane-Emden equation \citep[See][for an  
  in-depth description of polytropes]{Chandra1939}.  For the $n = 5$  
polytrope, the radial density distribution is of the same form as  
Equation \ref{EQN:PLUMMERRHO}.  Instead of setting a velocity field  
complimentary to the density field to support against collapse, a  
polytrope is supported by a thermal pressure gradient.    
The thermal energy of  
the gas is related to the velocity dispersion of the gas by equating  
it to the sound speed and then converting to specific internal energy  
by $u(r) = \sigma^2(r)/(\gamma - 1)$ where $\gamma$ is the ratio of  
specific heats of the gas.  We note that the gas itself in our  
simulation does not need to obey a polytropic equation of state, only  
that the thermal energy distribution of the gas is set-up to mimic the  
pressure distribution of the equilibrium polytrope and therefore  
remain in hydrostatic balance.  The gas responds adiabatically and  
therefore can heat by contraction or cool by expansion as it settles  
or is moved around by the potential of the stars.  The initial positions  
are set-up using the method of \citet{Aarseth1974}, but the thermal  
energies are set using the above equation and the initial velocities 
are set to zero.  
 
We simulate the evolution of (a) a gas-only $n = 5$ polytrope, (b) a  
star-only Plummer sphere, and (c) a 50-50 mixture (by mass) of a  
star-Plummer sphere and a $n = 5$ gas polytrope. 
Gas-only simulations are conducted with $N_g = 5,000$ gas particles 
of total mass $M_g$, and star-only simulations with $N_s = 500$ 
star particles of total mass $M_s$.  For each case, $M = M_s + M_g = 1$ 
and $a = 1$ using dimensionless units (where $G = 1$).   
Mixed star-gas simulations have either either $N_s = 100$ or $500$
star particles and either $10 \times$ or $100 \times$ as many gas particles.   
In all cases we use equal-mass star particles ($m_s = M_s / N_s$) and 
equal-mass gas particles 
($m_g = M_g / N_g$)\footnote{Equal-mass star particles is obviously a 
  simplification for the purposes of our tests.  However equal-mass gas 
  particles should be used to reduce SPH noise and prevent particle 
  clumping.}. 
The smoothing length of the stars in all simulations is $h = 0.0001\,a$.
Each simulation is run for 40 crossing times, where we define the  
crossing time here to be $t_{_{\rm CR}} = a/\sigma(r = 0) =  
\sqrt{6\,a^3\,/\,G\,M} = 2.45$ code time units.

\begin{figure*}  
\centerline{\psfig{figure=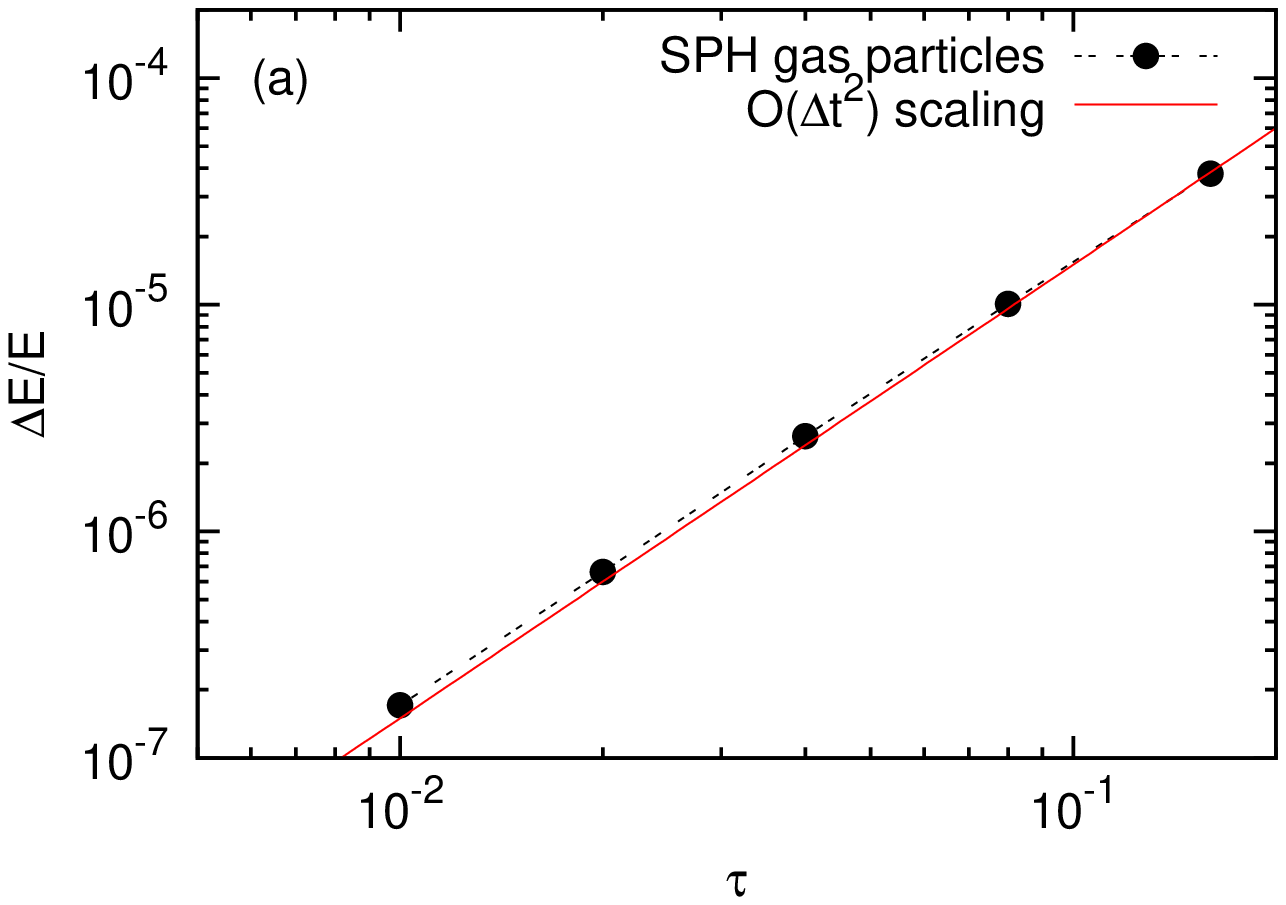,width=8.0cm,angle=270}  
\hspace{0.02cm}\psfig{figure=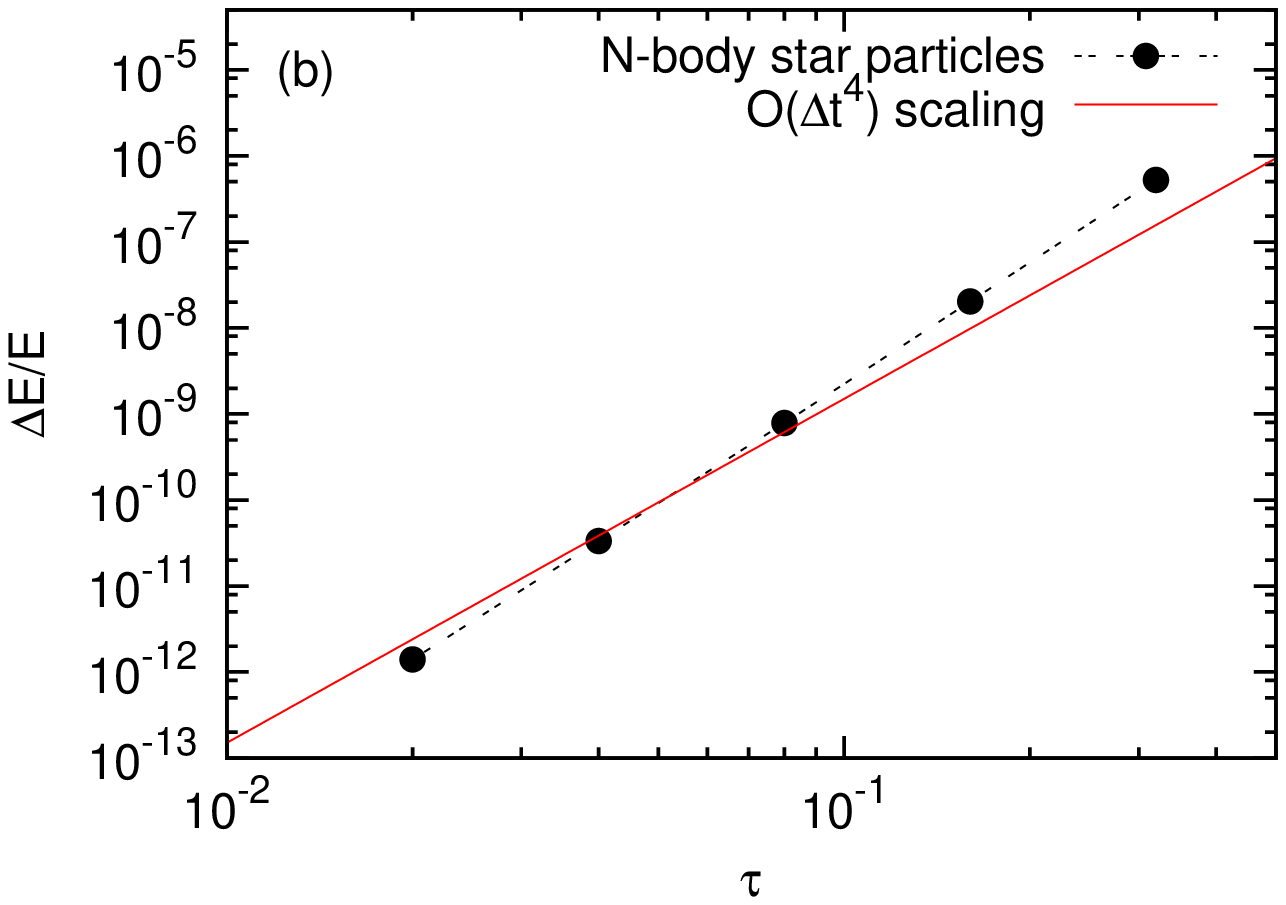,width=8.0cm,angle=270}}  
\centerline{\psfig{figure=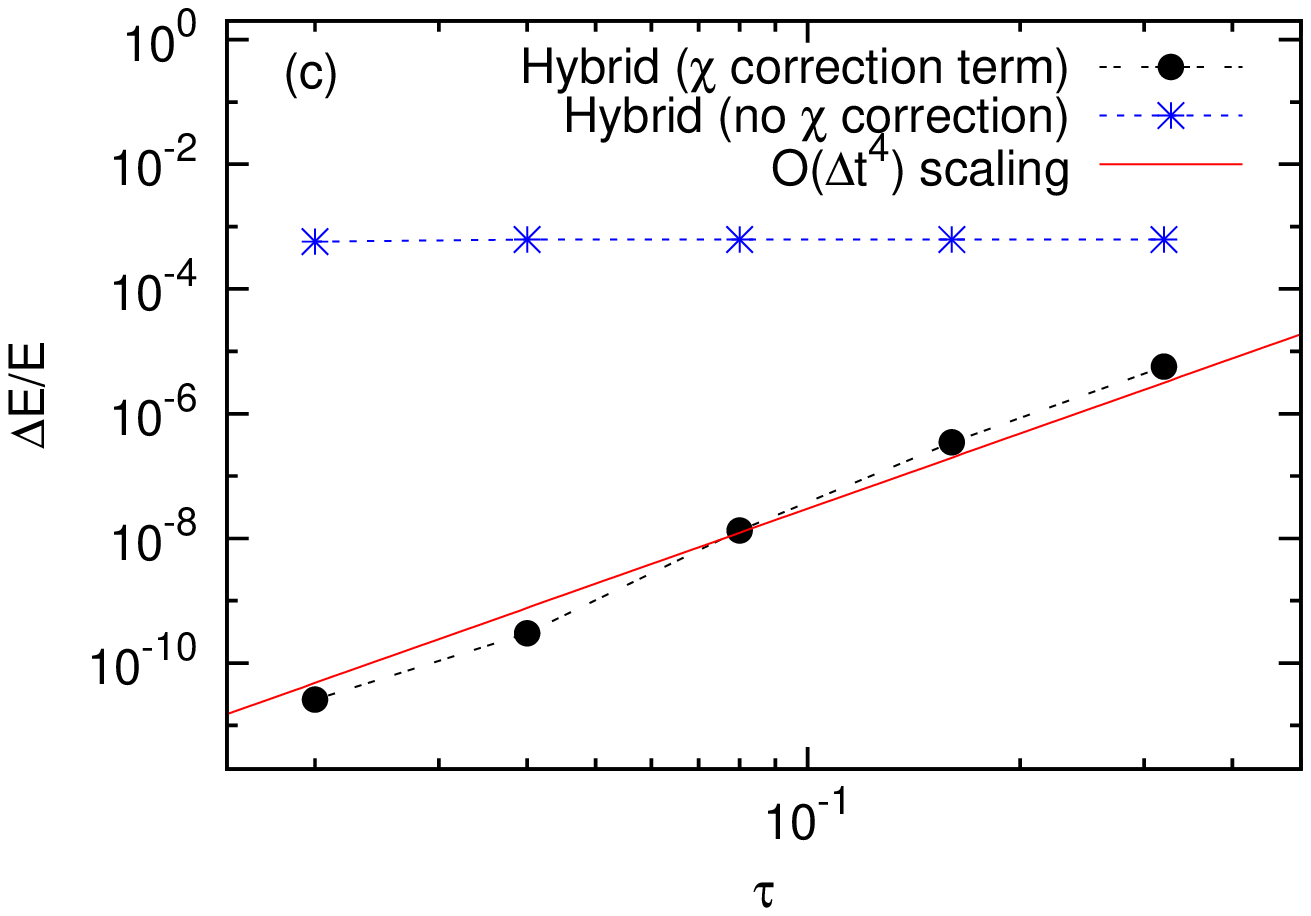,width=8.1cm,angle=270}}  
\caption{The fractional global energy error, $\Delta E/E$, as a function of timestep multiplier, $\tau$ for (a) SPH gas particles in a $n = 5$ polytrope integrated with a 2nd-order Leapfrog, (b) $N$-body star particles in a Plummer sphere integrated with a 4th-order Hermite, and (c) SPH gas and star particles in a Plummer/Polytrope combination where the gas is integrated with a 2nd-order Leapfrog and the stars with a 4th-order Hermite, with and without the $\bar{\chi}$ correction term derived in Section \ref{SS:GRADHSPH} and Appendix \ref{A:SPHDERIVATION}.  Also plotted for guidance are the 2nd- and 4th-order scaling expected for the SPH, $N$-body { and hybrid} simulations.}
\label{FIG:ENERGYERROR}  
\end{figure*}  

\subsubsection{Resolution criteria}  \label{SSS:PLUMMERRESOLUTION}
  
Following the ideas discussed in Section \ref{SS:SCATTERINGTEST}, we can determine the resolution requirements of an equilibrium Plummer sphere to significantly reduce the effects of unphysical star-gas scattering.  Let us assume that $N_g$ gas particles account for a fraction $f$ of the total mass, i.e. $M_g = f\,M = N_g\,m_g$ where each gas particle has mass $m_g$.   The central gas density is $\rho_0 = 3\,M_g/4\,\pi\,a^3$, and the central velocity dispersion is $\sigma^2_0 = G\,M/ 6\,a$.  We can then calculate the critical resolution velocity at the centre of the Plummer sphere as   
\begin{eqnarray} \label{EQN:VCRITPLUMMER}  
v_{_{\rm CRIT}} &=& \left(\frac{2\,G\,m_g}{h} \right)^{1/2} = \left(\frac{2\,f\,G\,M}{\eta\,a}\right)^{1/2} \left(\frac{3}{4\,\pi} \right)^{1/3}\, N^{-1/3}  
\end{eqnarray}  
where we have substituted Equation \ref{EQN:HRHO} for $h$ and used the above expressions for $\rho_0$.
  
In order to avoid catastrophic numerical scattering of star particles off gas particles, star particles must be moving at velocities significantly larger than the critical velocity, i.e. $\sigma_0 \gg v_{\rm{c}}$.   This leads to the following resolution criterion for the total gas particle number, $N$, in the Plummer sphere as  
\begin{eqnarray} \label{EQN:PLUMMERRESOLUTION}  
N &\gg& \left(\frac{12\,f}{\eta} \right)^{3/2} \left(\frac{3}{4 \pi} \right)^{1/2} \approx 15\,f^{3/2} 
\end{eqnarray} 
assuming the typical value of $\eta = 1.2$.   
For a Plummer sphere consisting of equal gas and stellar mass ($f =
0.5$) we find that $N_g \gg 5$.  For a gas-dominated Plummer sphere
($f = 1.0$), $N_g \gg 15$.  \citet{Hubber2011} demonstrated of order a 
hundred gas particles could only crudely represent an equilibrium
polytrope, with reduced central density and a smaller radius.  Such
structures require of order thousands or tens of thousands of gas particles to
adequately resolve the density structure of the polytrope.  Therefore,
it is clear that we require $N \gtrsim 1000$ to resolve the density field, at which point the gas resolution is also sufficiently high to prevent serious star-gas scattering events from corrupting the simulation.  

 We note that the stars are moving with a range of velocities, below and above the mean velocity dispersion, $\sigma_0$.  No matter how high the resolution, there will always be a number of stars at some instant moving less than the critical velocity.  Therefore, we cannot completely eliminate unphysical scattering in this case, but we can only reduce it to some acceptable level by using a reasonably high resolution.
  
\begin{figure*}  
\centerline{\psfig{figure=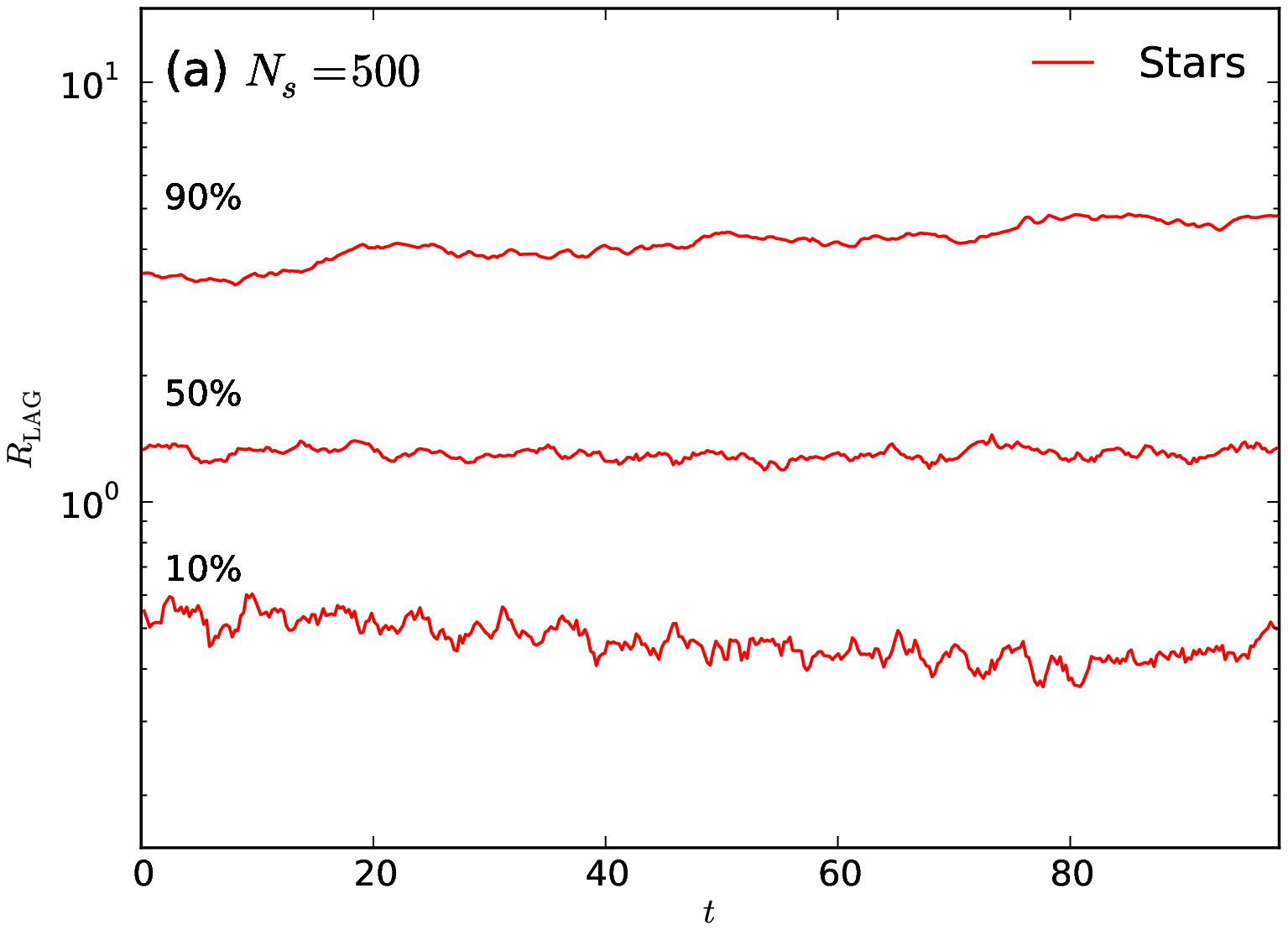,width=8.5cm,angle=0}  
\hspace{0.02cm}\psfig{figure=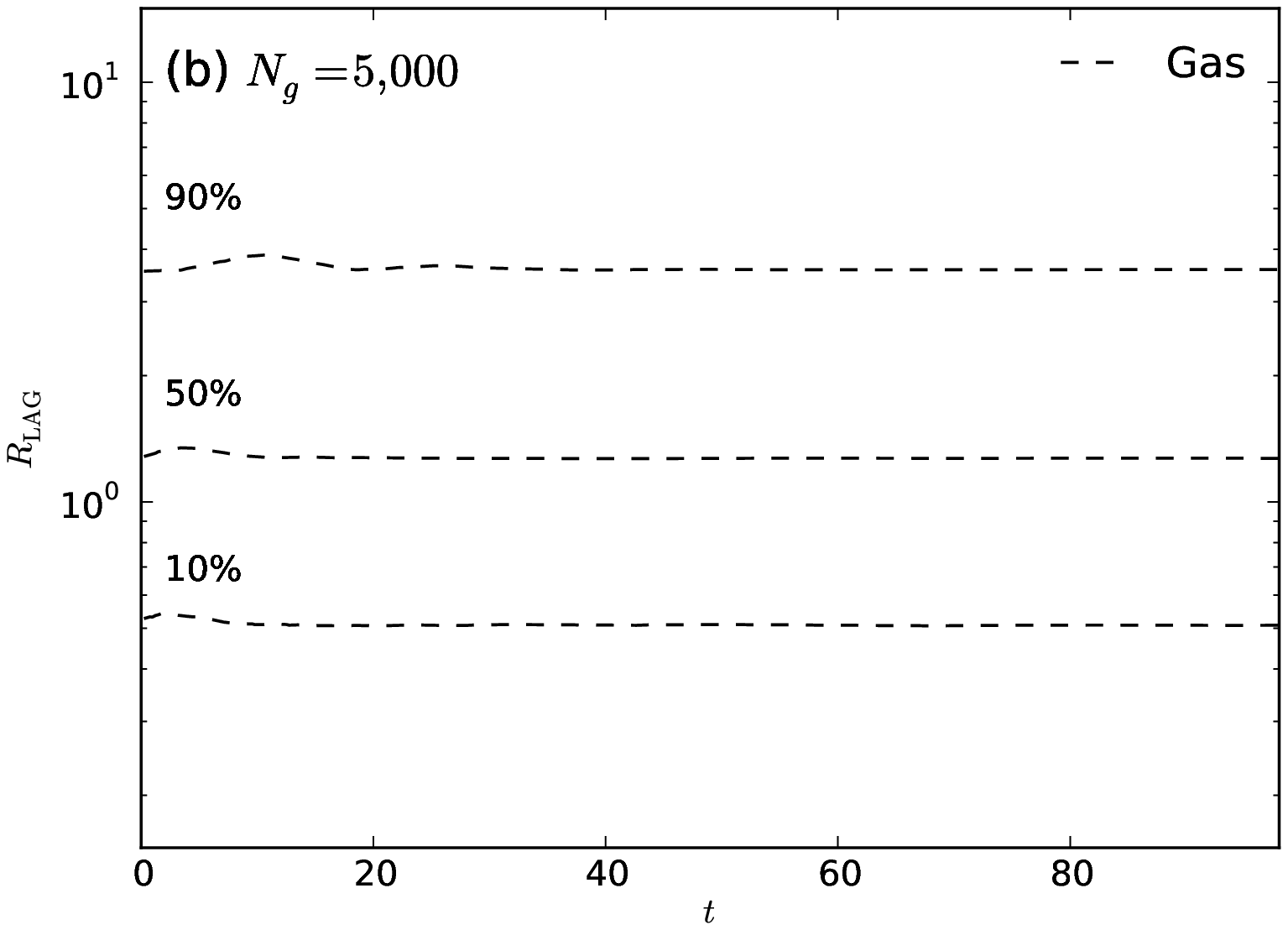,width=8.5cm,angle=0}}  
\centerline{\psfig{figure=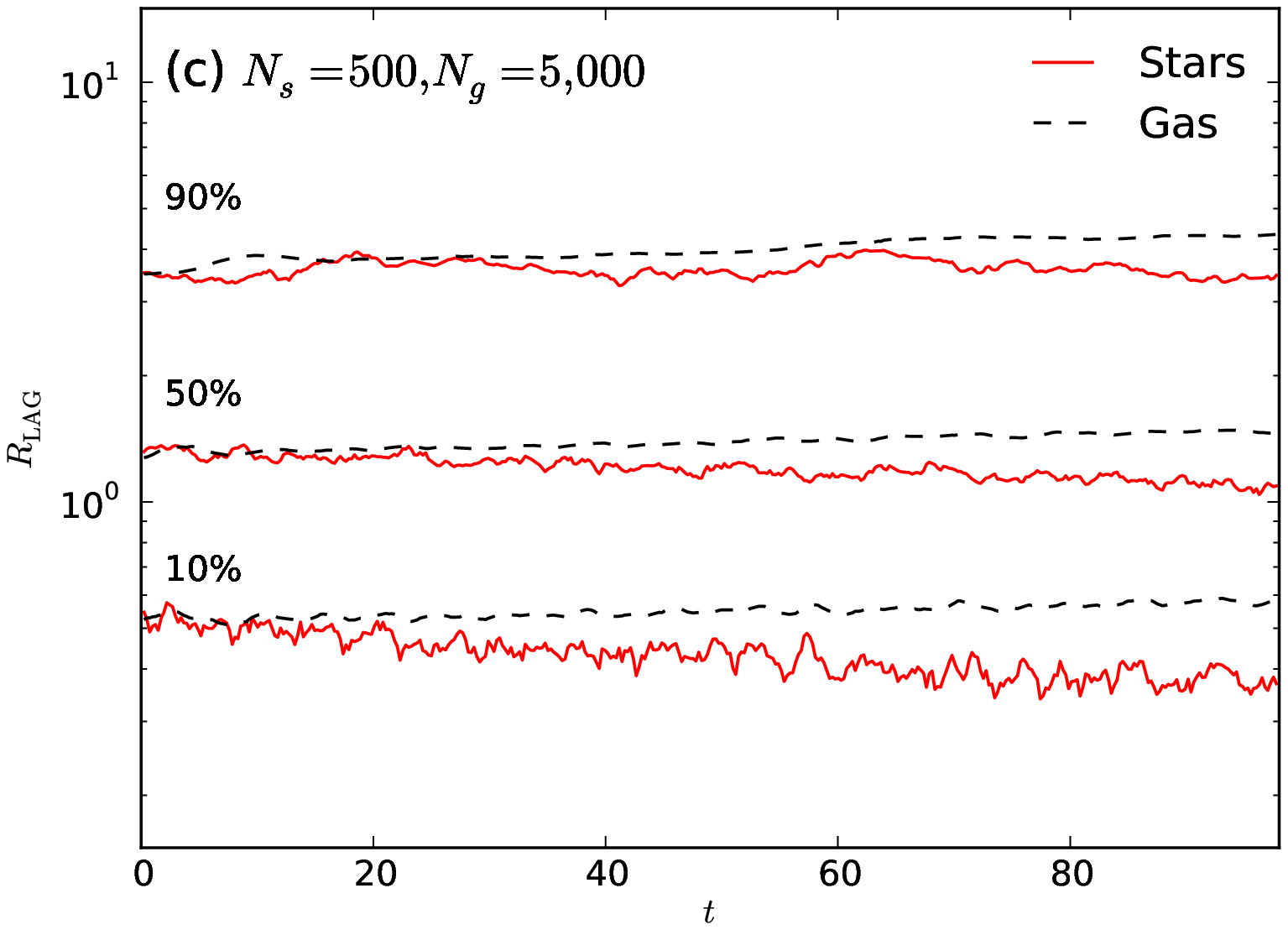,width=8.5cm,angle=0}  
\hspace{0.02cm}\psfig{figure=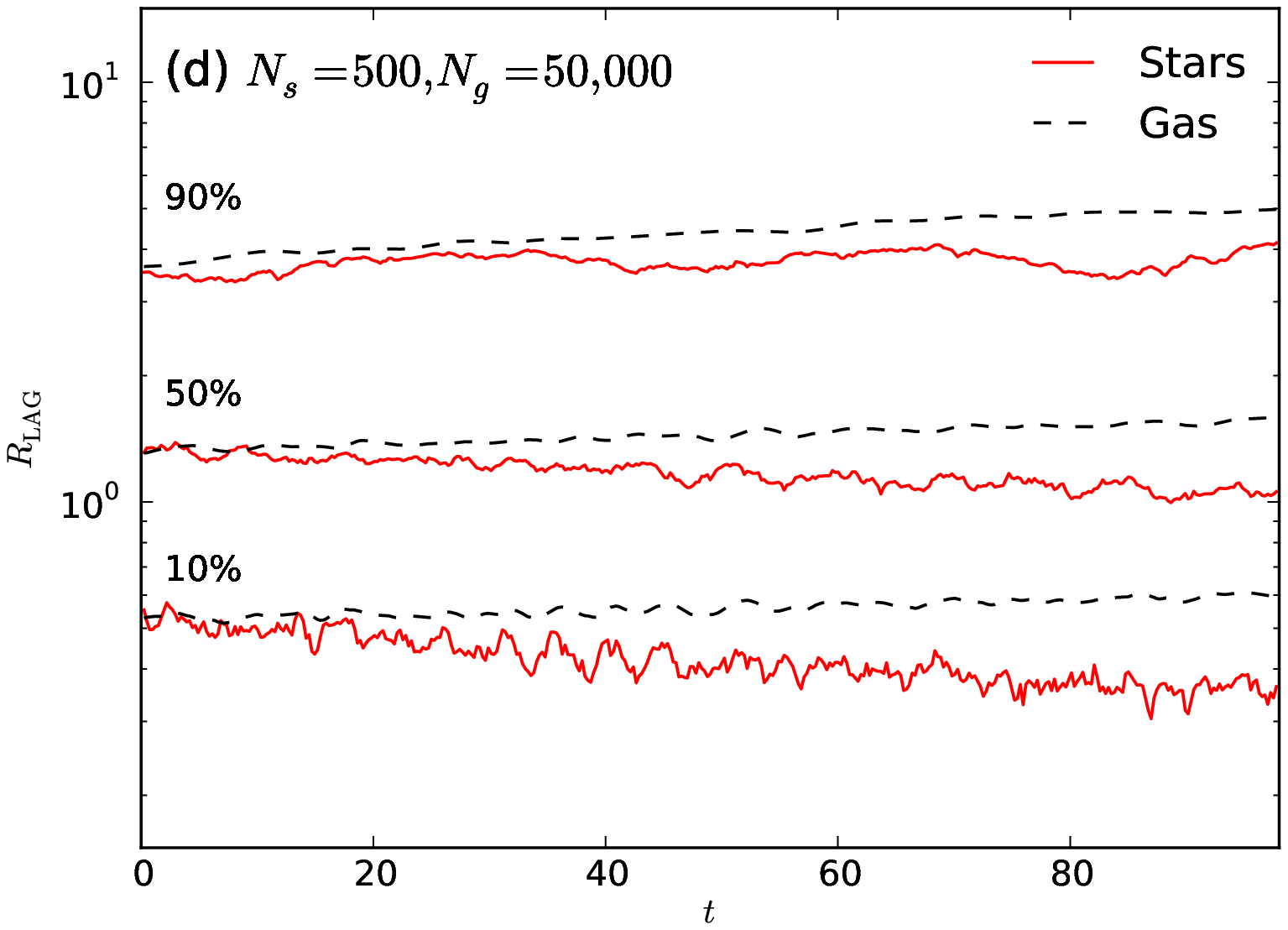,width=8.5cm,angle=0}}  
\centerline{\psfig{figure=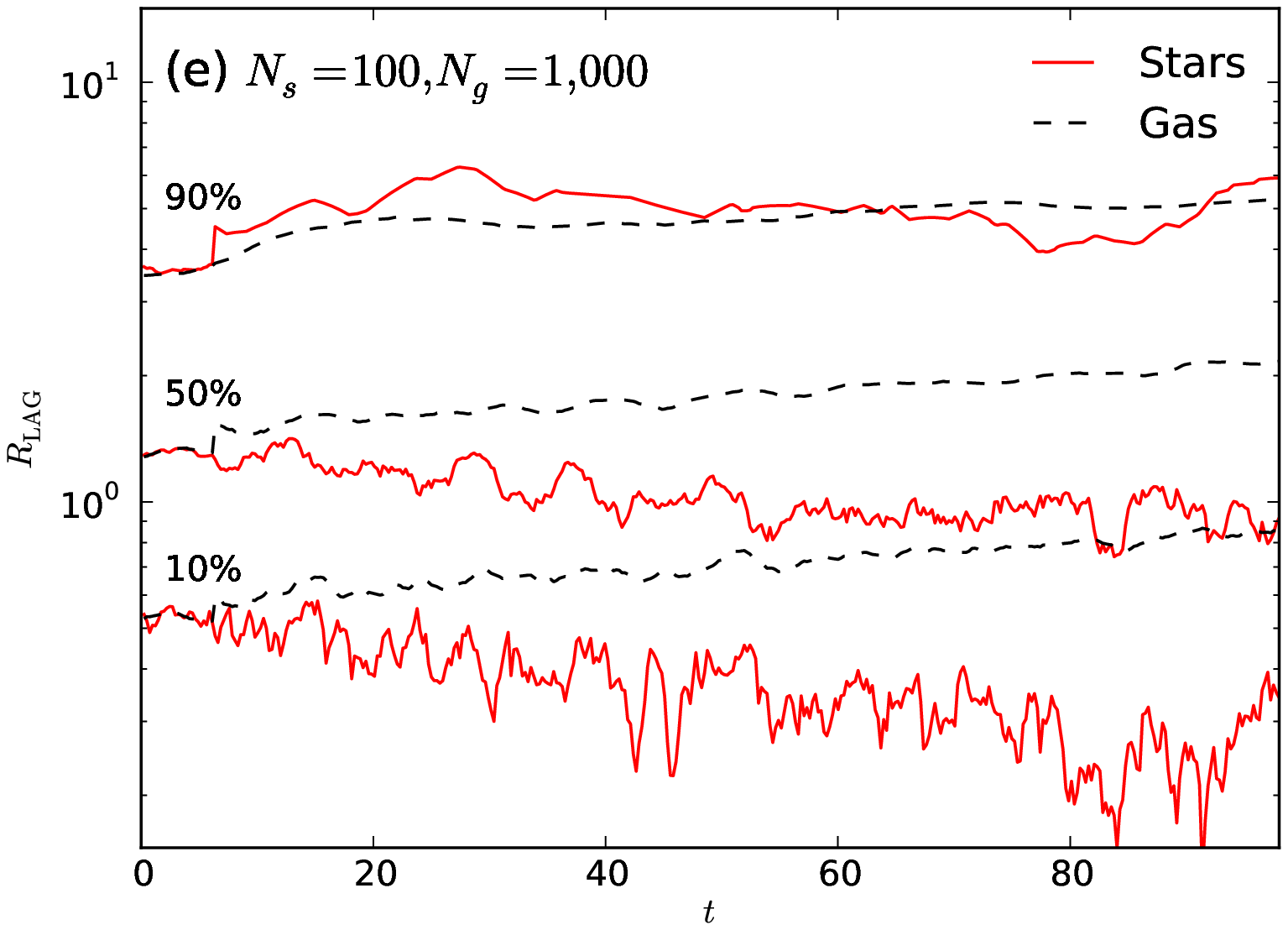,width=8.5cm,angle=0}  
\hspace{0.02cm}\psfig{figure=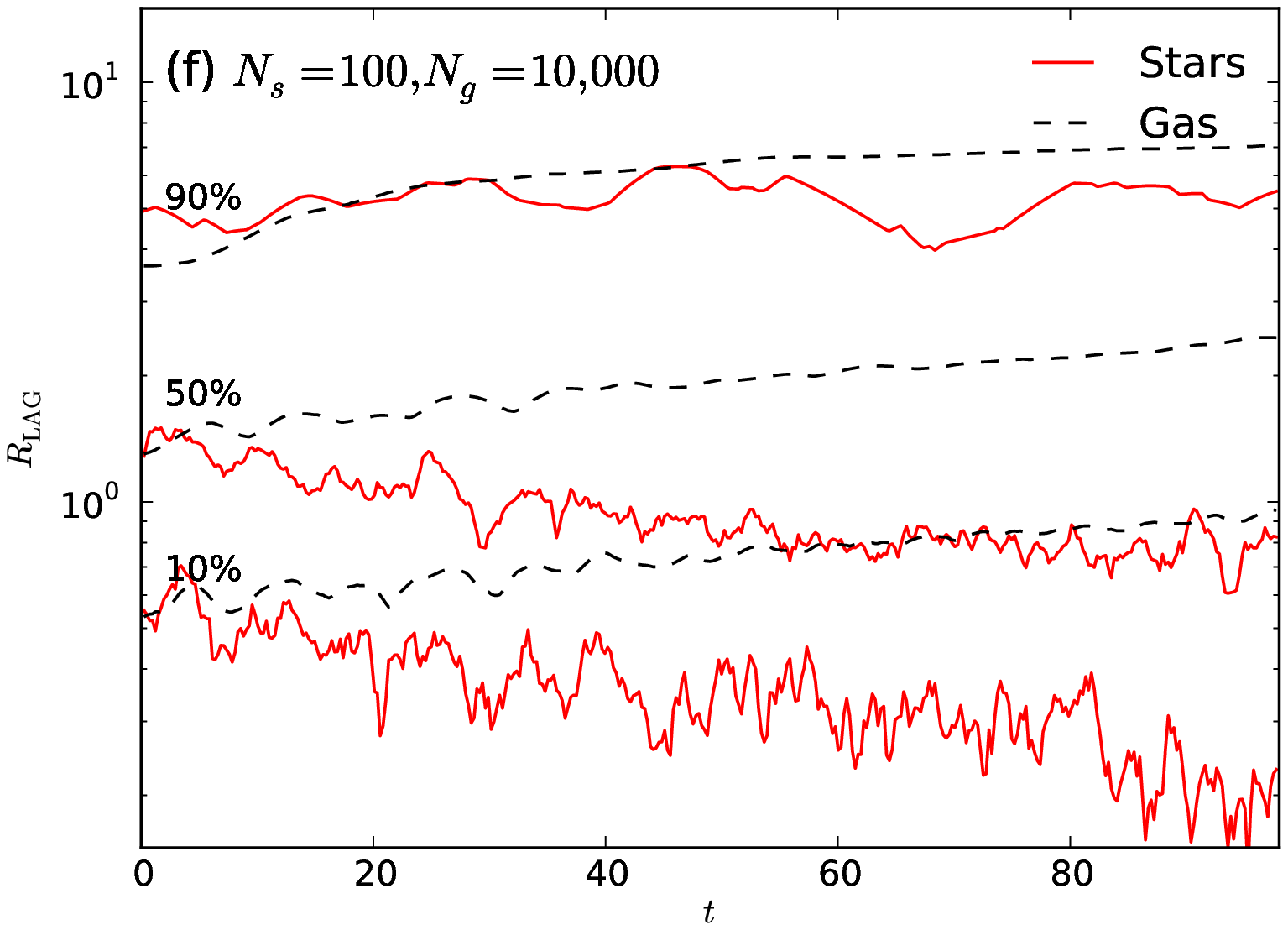,width=8.5cm,angle=0}}  
\caption{The time evolution of the $10\%$, $50\%$ and $90\%$ Lagrangian radii for Plummer spheres of $M = 1$, $a = 1$ (in dimensionless units) containing (a) $500$ stars only, (b) $5,000$ gas particles only, (c) $5,000$ gas particles and $500$ stars, (d) $50,000$ gas particles and $500$ stars, (e) $1,000$ gas particles and $100$ stars, and (f) $10,000$ gas particles and $100$ stars.}  
\label{FIG:LAGRANGIANRADII}  
\end{figure*}

\subsubsection{Energy conservation} \label{SSS:ENERGYTEST}  
  
In order to test the energy conservation properties of hybrid code, we investigate how the fractional global energy error, $\Delta E/E$, varies as a function of the timestep size.  Instead of selecting a global, constant timestep, we allow an adaptive global timestep, and then vary the timestep multiplier, $\tau$ (to which we equate all other timestep multipliers, $\gamma_{_{\rm COUR}}$, $\gamma_{_{\rm ACCEL}}$, $\gamma_{_{\rm ENERGY}}$ and $\gamma_{s}$, for this test), which determines the adaptive timestep size.   Therefore we consider how the energy error scales with $\tau$, which should scale in the same way with a constant stepsize.  
  
Figure \ref{FIG:ENERGYERROR} shows the energy error scaling as a  
function of timestep multiplier for (a) gas-only polytrope, (b)  
star-only Plummer sphere, and (c) the star-gas mixture as a function  
of $\tau$.  For the gas-only cluster, the SPH integration scheme is a  
2nd-order Leapfrog kick-drift-kick.  Therefore we expect the error to  
scale as $|\Delta E| \sim {\cal O}(dt^2) \sim {\cal O}(\tau^2)$.  In  
Figure \ref{FIG:ENERGYERROR}(a), we can see that the energy error  
(filled black circles) agrees very well with this expected scaling  
with only a small deviation from the added guideline (solid red line).  
For the star-only cluster, the $N$-body integration scheme is a  
4th-order Leapfrog scheme, and therefore we expect $|\Delta E| \sim  
{\cal O}(dt^4) \sim {\cal O}(\tau^4)$.  Figure  
\ref{FIG:ENERGYERROR}(b) shows that we obtain similar scaling to this.  
In fact, we obtain slightly steeper scaling relative to the expected  
4th-order.  
{ For the star-gas cluster, a combination of a 2nd-order  
scheme with a 4th-order scheme should in principle give either 2nd-order  
 or 4th-order erros depending on whether gas-gas, star-gas or star-star 
interactions are dominating the error.  Figure \ref{FIG:ENERGYERROR}(c) 
shows that we get 4th-order scaling (filled black circles) which 
suggests that star integration scheme dominates the error in this simple case, 
despite the formally larger error of the gas integration scheme.}  
We also plot the energy errors for they 
star-gas cluster with and without the new $\bar{\chi}$ correction term 
(Equations \ref{EQN:SPHEOM} \& \ref{EQN:GRADHCHI}) introduced in this paper.  
Without $\bar{\chi}$, the equations of motion are non-conservative, and the 
energy error (of order $10^{-3}$) is dominated by this rather than 
integration error. 

As well as integration error, block timesteps and gravity tree errors are other major sources of error in this scheme, and will likely dominate over the integration error in most practical simulations depending on the parameters chosen.  The error in the tree can be controlled by an appropriate choice of multipole-acceptance criteria (MAC), such as the GADGET-style MAC \citep{Gadget1} or the SEREN Eigenvalue MAC \citep{Hubber2011}, both of which can place an upper-bound on the force error due to individual cells and hence control the net tree force error and indirectly the global energy error.

\subsubsection{Plummer sphere stability} \label{SSS:STABILITYTEST} 
  
One of the aims of this hybrid scheme is to accurately follow the stellar  
dynamics of a system in a live gaseous background.  The most  
significant difference between stars and gas in this context is  
that stars are collisional particles in the sense that they can be  
subject to strong two-body interactions, whilst gas is collisional  
in the sense that it can form shocks.

A low-$N$ stellar Plummer sphere can rapidly evolve due to two-body  
scattering.  \citet{Aarseth1974} showed, both numerically and through  
semi-analytical models, that the effect of two-body  
relaxation in a Plummer sphere is to redistribute energy ejecting some  
stars and causing the contraction (eventually to core collapse) of the  
remaining cluster.  \citet{Aarseth1974} traced this with the Lagrangian  
radii of the stellar clusters showing the contraction of the inner  
Lagrangian radii and the expansion of the outer Lagrangian radii due  
to energy conservation as stars are ejected.  They showed that the 50  
per cent Lagrangian radius stays relatively constant throughout the  
evolution.  The Lagrangian radii evolve significantly on the two-body  
relaxation timescale, $t_{\rm relax}$, given in terms of the crossing  
time, $t_{\rm cross}$, as  
\begin{eqnarray} \label{EQN:TRELAX}  
t_{\rm relax} &\sim& \frac{0.1\,N_s}{\ln{N_s}}\,t_{\rm cross}  
\end{eqnarray}  
where $N_s$ is the number of stars \citep[See][]{GalacticDynamics2}. 
  
The evolution of the star-only Plummer sphere (Figure
\ref{FIG:LAGRANGIANRADII}(a)), containing $500$ stars, shows the same
qualitative behaviour as found by \citet{Aarseth1974} demonstrating
that our pure $N$-body integration scheme is correctly capturing the
qualitative effect of 2-body encounters over the expected timescale
(for $N_g = 500$, the relaxation timescale is $t_{\rm relax} \sim
10\,t_{\rm cross}$ using Eqn. \ref{EQN:TRELAX}).  As observed by Aarseth
(1974) and Aarseth et al. (1974), the 10 per cent Lagrangian radius
shrinks (towards core collapse), the 90 per cent Lagrangian radii
expands (due to ejections), and the 50 per cent Lagrangian radii stays
roughly constant.

The gas-only Plummer sphere (Figure \ref{FIG:LAGRANGIANRADII}(b)),
containing $5,000$ gas particles, is observed to evolve slightly over
the first few crossing times as it settles into equilibrium but soon
the Lagrangian radii stay almost constant over time.
  
Of far greater interest is the behaviour of a mixed star-gas Plummer  
sphere.  We run two sets of simulations: $100$ star particles with
either $1 000$ or $10 000$ gas particles, and $500$ star particles
with either $5 000$ or $50 0000$ gas particles.  We use two different 
gas resolutions for each case to help determine if the simulations are 
converged.  
The evolution of the 10, 50 and 90 per cent Lagrangian radii of both the 
stars and gas in each case are shown in Figure \ref{FIG:LAGRANGIANRADII}. 
  
We notice, for both $N_s = 500$ and $N_s = 100$, that the Lagrangian radii for the stellar and gas components evolve in opposite directions.  The stellar component is altered somewhat from the star-only case where it appears to shrink for the 10 per cent and 50 per cent Lagrangian radii (with the 90 per cent remaining fairly constant).  Conversely, the gas appears to expand at all radii, and on a timescale comparable to the stellar 2-body relaxation timescale.  Therefore, there is a transfer of energy from the stars to the gas, allowing the gas to heat and expand, and conversely the stars lose energy and contract.

Most importantly for this paper, the results converge for different
gas particle numbers (with the same number of star particles).  For
$N_s = 500$, we use both $5,000$ (Figure \ref{FIG:LAGRANGIANRADII}(c))
and $50,000$ (Figure \ref{FIG:LAGRANGIANRADII}(d)) gas particles. The
evolution of both the $N_s = 500$ simulations is basically identical.  
There is some deviation between the $N_s = 100$ results at late times
in different $N_g$ backgrounds.  This is due to low-$N_s$ noise
and the slightly earlier `core collapse' of the $N_g = 10,000$
simulation.  

The reader might notice that the behaviour of the stars in the $N_s =
500$ star-only simulation is somewhat different to that of the stars
in the $N_s = 500$ simulations with gas.  This is an interesting
physical (not numerical) effect due to the presence of gas.  We will
return to the physics and astrophysical implications of this behaviour
in the next paper.  For now, however, we will simply use these
simulations to illustrate the convergence of the results for 
different numbers of gas particles but the same number of star particles.

\begin{figure*}  
\centerline{\psfig{figure=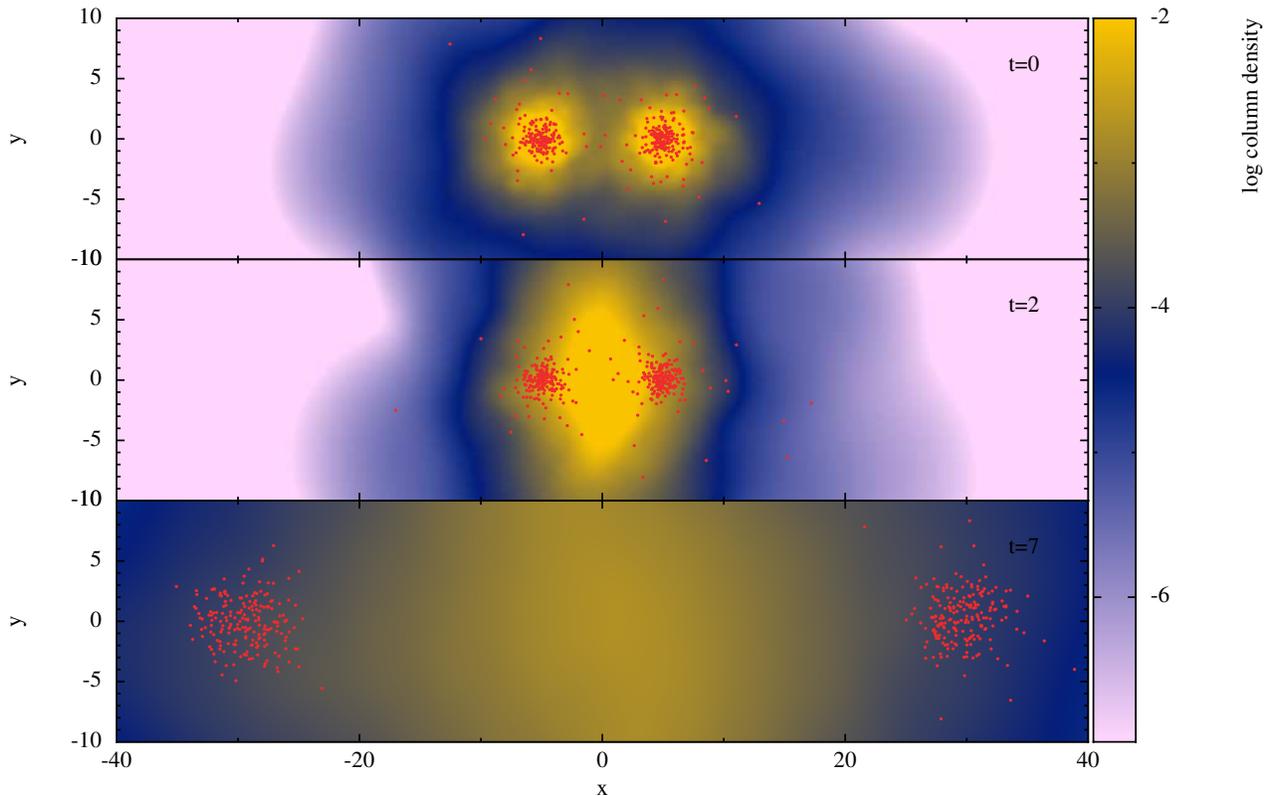,width=17.0cm,angle=0}}  
\caption{Supersonic collisions between gas-dominated (90 per cent gas
  by mass) Plummer spheres at (a) the initial state; (b) just after the  
collision; (c) the end of the simulation.Each Plummer sphere has
$5000$ gas particles and $200$ equal-mass star particles.  The initial 
crossing time of a Plummer sphere is 2.45 code units and the time is
measured in code units.  Stars are shown by white dots, the colour
table shows the column density of the gas in code units.}  
\label{FIG:CP-GHSS}  
\end{figure*}

\begin{figure*}  
\centerline{\psfig{figure=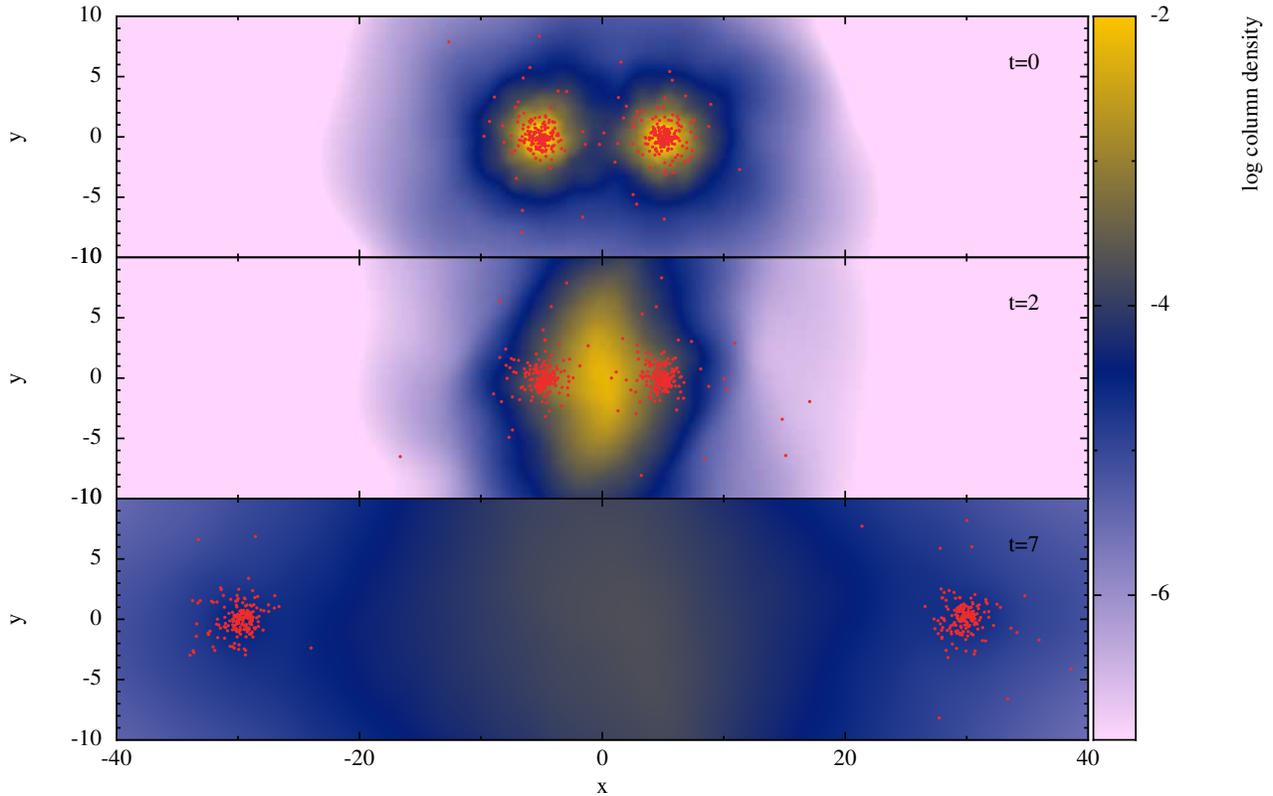,width=17.0cm,angle=0}}  
\caption{Supersonic collisions between star-dominated (90 per cent
  stars by mass) Plummer spheres.  Otherwise the same as fig.~\ref{FIG:CP-GHSS}.}  
\label{FIG:CP-SHSS}  
\end{figure*}

\subsection{Star-Gas cluster collisions} \label{SS:COLLISIONTEST}  
  
As a simple qualitative test of the hybrid code's ability to model more complex star-gas systems, we perform a small suite of simulations of the head-on impact between two star-gas Plummer spheres.  We create two star-gas Plummer spheres following the procedure described in Section \ref{SS:PLUMMERTEST}.  We collide the Plummer spheres at a velocity $v_{\rm{coll}}$, such that the collision occurs either subsonically or supersonically for all gas particles.  We also consider Plummer spheres that are gas-dominated, and star-dominated.  We therefore expect strong differences in the behaviour of the gas and stellar dynamics between the subsonic and supersonic tests, and also between the star and gas-domainted cases.
 
Each Plummer sphere contains $N_g = 5,000$ equal-mass gas particles and $N_s = 200$ equal-mass star particles and is set-up in the same way as in Section \ref{SS:PLUMMERTEST}.  For gas-dominated cases, 90 per cent of the mass is in gas and for star-dominated cases, 90 per cent of the mass is in stars (therefore the relative masses of star and gas particles are different by a factor of $\sim 100$ between the two cases).  For subsonic collisions, the relative collision velocity is $0.5$ (in dimensionless units), and for supersonic collisions, the relative collision velocity is $10$.

Figures \ref{FIG:CP-GHSS} and \ref{FIG:CP-SHSS} show the results of supersonic collisions of a gas-dominated and a star-dominated collision respectively.  In the gas-dominated supersonic collision (Fig. \ref{FIG:CP-GHSS}), the gas forms a shock around $x = 0$ where the gas is heated up and is compressed to higher densities.  Meanwhile the stars pass through the shock front and also through the stellar-component of the other cluster.  Since the relative velocity of the two clusters is much greater than the individual velocity dispersions, the effects of two-body encounters are negligible and the two clusters pass through each other almost unperturbed. As the gaseous and stellar components have decoupled in the collision, the gas-free stellar clusters are now unbound as they have had 90 per cent of their initial mass (the gas) removed.  Therefore they expand as their velocity dispersion is too high to maintain their initial dense configuration; the cluster eventually dissolves over several crossing times  (this is analogous to gas expulsion, see \citealt{GoodwinBastian2006}).  In the star-dominated supersonic collision (Fig. \ref{FIG:CP-SHSS}), the gas again shocks and decouples from the stars.  However, as the stars dominate the potential, the cluster is only slightly super-virial ($\frac{1}{2} < Q < 1$) and the degree of expansion whilst readjusting to the new potential is small.  Therefore, the two stellar clusters continue with almost no effect from the collision and the stripping of their gas. 
  
\begin{figure*}  
\centerline{\psfig{figure=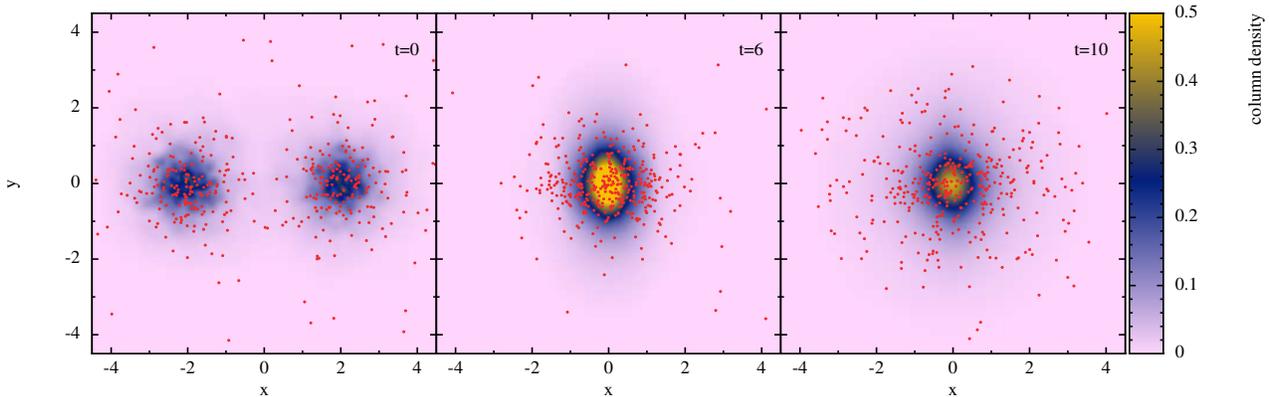,width=17.0cm,angle=0}}  
\caption{Subsonic collisions between gas-dominated (90 per cent gas by mass) Plummer spheres.  Otherwise the same as fig.~\ref{FIG:CP-GHSS}.}
\label{FIG:CP-GHNS}  
\end{figure*}

\begin{figure*}  
\centerline{\psfig{figure=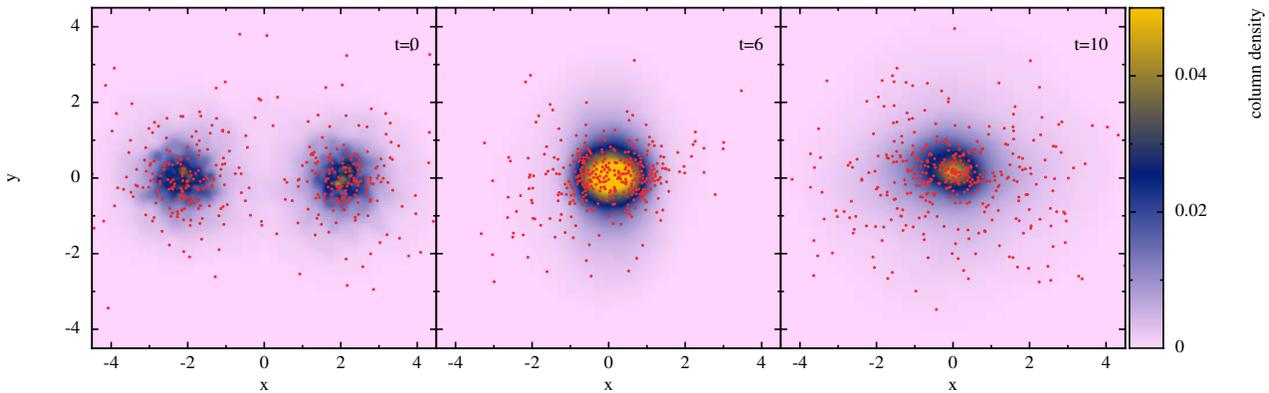,width=17.0cm,angle=0}}  
\caption{Subsonic collisions between star-dominated (90 per cent stars by mass) Plummer spheres.  Otherwise the same as fig.~\ref{FIG:CP-GHSS}.}
\label{FIG:CP-SHNS}  
\end{figure*}

In Figures \ref{FIG:CP-GHNS} and \ref{FIG:CP-SHNS}, we show the results of subsonic collisions of gas-dominated and star-dominated clusters respectively.  In both cases the two clusters merge as the gas components merge and the stellar components can respond to the interaction as they interact on a timescale comparable to their crossing times.  The shapes and details of the resulting clusters are different; the star-dominated merger showing a more elongated final appearance (compare the last panels of Figures \ref{FIG:CP-GHNS} and \ref{FIG:CP-SHNS}).  This is explained as the post-shock velocity anisotropy of the stellar velocity field dominates in the star-dominated case, whilst in the gas-dominated merger the potential is dominated by the gas allowing significant violent relaxation of the stellar component in the spherical gas potential.
  
The behaviour of the supersonic and subsonic collisions in these simulations is physically reasonable.  We also track energy conservation and find it is very well conserved ({ $\Delta E/E = 1.6 \times 10^{-6} - 1.4 \times 10^{-5}$ for the four simulations}) .  As with the static Plummer test, we will explore the physics of star-gas cluster collisions in more detail in a future paper.

\section{Discussion} \label{S:DISCUSSION}  
  
Our main motivation for developing this new hybrid method is
to provide an intermediate step between detailed hydrodynamical 
simulations and pure
$N$-body simulations.  In particular, we wish to investigate the
dynamics of stars in gas on $>0.1$~pc to pc-scales in GMCs and young
clusters.  The main advantage of our method is that the dynamics of
stars within a live gas background can be followed at relatively  
low computational expense compared to full hydrodynamical
simulations (hours or days on desktop computers compared to months on
HPC facilities).  We discuss a number of important practical considerations 
for preparing simulations using the hybrid code, as well as planned 
future developments and uses of the code.

\subsection{Suppressing fragmentation}

We reiterate that this method is not intended to act as a replacement for full hydrodynamical simulations which aim to model the star formation process itself, i.e. the fragmentation of molecular clouds into prestellar cores and finally multiple protostellar systems.  The fragmentation of gas into stars is a complex hydrodynamical problem, involving much additional physics such as radiation transport, chemistry, and (non-ideal) MHD.   We suggest that fragmentation should be artificially supressed for a number of reasons.  Firstly, we wish to avoid the complex physics and computational expense related to full hydrodynamical simulations since we are only currently interested in the global effects of the gaseous gravitational potential on the $N$-body evolution of the cluster.  Secondly, if fragmentation occurs then sink particles should be introduced and it is unclear how to mix sink particles with $N$-body star particles (one will be softened and interacting hydrodynamically with the gas, the other will not despite both representing a star).  

Therefore we suggest that the resolution be kept {\em as low as possible} and that the equation of state be designed to supress star formation and keep densities low.  We note that this also has the advantage of avoiding the formation of discs around stars which are again complex, computationally expensive objects to model.  We suggest that hybrid simulations are designed to avoid the regimes in which these processes are important, ie. sub-core ($<0.1$~pc) scales and high gas densities ($>10^{-13}$~g~cm$^{-3}$).
 
The normal resolution criteria for SPH simulations of star formation is the \citet{BateBurkert} criteria in which the (minimum) Jeans mass must be resolved in order to capture gravitational fragmentation \citep[or in AMR, the][Jeans length criterion]{Truelove1997}.  As shown by \citet{Hubber2006}, failure to meet the \citet{BateBurkert} criteria means that fragmentation is suppressed -- this is important as it means that low SPH resolution results in no fragmentation rather than artificial fragmentation.  
 
We suggest that gas be kept at densities lower than 
$10^{-20} - 10^{-15}$ g cm$^{-3}$, well above the critical density 
for fragmentation \citep[i.e. $10^{-13}$ g cm$^{-3}$,][]{MI2000}.  
Such densities are in the roughly isothermal regime, 
and so have a simple equation of
state \citep[e.g.][]{Jappsen2005}.  We suggest that the equation of state be
modified (say by artificial heating) to keep densities low.  The
potential to form cores is probably desirable, but not to follow 
their collapse and fragmentation.

\subsection{Avoiding unphysical scattering}

As discussed and demonstrated in the star-gas scattering test  (Section \ref{SS:SCATTERINGTEST}), a sufficiently high gas resolution is required to avoid the unphysical scattering of stars by gas particles.  For a general situation, where there is a group of stars moving through a cloud of gas, but no equilbrium has been established, then the velocities are not linked in any way to the density of the gas and it is difficult to establish a simple criterion for the required gas resolution.  There are various scenarios where we can establish a link between the star velocity and the gas density.  For example, equilibrium Plummer spheres, we were able to derive a resolution condition for equilibrium Plummer spheres on the number of gas particles which was only a function of the gas mass fraction, $f$ (Eqn. \ref{EQN:PLUMMERRESOLUTION}).  We note that this condition is only strictly true for equilibrium clusters where the velocity (or velocity dispersion) of the stars is well-known.  For non-equilbrium simulations, we may need to use further information infered from the initial conditions, such as the initial virial ratio, to infer how the velocity relates to the density, and hence to the required resolution of the gas.  

One other special scenario is a gaseous cluster with emedded primoridal binary and multiple systems which may be unphysically disrupted due to star-gas scattering. 
Consider the simple case of a binary star containing two stars of mass $m_1$ and $m_2$ in a circular orbit of separation $a_{_{\rm BIN}}$, with an orbital velocity  $v_{_{\rm BIN}} = \sqrt{G (m_1 + m_2)/a_{_{\rm BIN}}}$.    
To avoid artificial scattering from potential destroying the binary, we require that $v_{_{\rm BIN}} \gg v_{_{\rm CRIT}}$.  Assume that a binary is located at the centre of a pure gas plummer sphere, consisting of $N$ gas particles, of total mass $M$, and scale-length $a$. The critical velocity of the central gas is given by Equation \ref{EQN:VCRITPLUMMER}.  Rearranging leads to the resolution condition,  
\begin{eqnarray} \label{EQN:BINARYRESOLUTION}  
N &\gg& \left( \frac{2\,f}{\eta} \right)^{3/2} \, \left(\frac{3}{4\pi} \right)^{1/2}\,\left[\frac{M}{(m_1 + m_2)} \frac{a_{_{\rm BIN}}}{a}\right]^{3/2} \,.  
\end{eqnarray}  
Inserting in reasonable values of $a=0.1$~pc, $a_{_{\rm BIN}}=10$~AU, $M=10^4~M_\odot$, and $m_1+m_2=0.1~M_\odot$, we find $N \gg 200$. Therefore, for a smooth distribution of gas, $N \gtrsim 10,000$ is sufficient to avoid star-gas particle scattering.  We note that wide binaries would be most sensitive to star-gas scattering in comparison to tight binaries.  Therefore, using wider-separation primordial binaries would require higher resolution.  However this scenario is highly idealised compared to more practical scenarios.  A more irregular gas distribution may have higher values of $v_{\rm{crit}}$ in high density pockets resulting in stronger scattering and therefore, more stringent resolution requirements.

\subsection{Future applications and code developments} 
  
There are many possible future astrophysical applications for our
hybrid method.  Our first follow-up papers will address stellar
dynamics in small-$N$ star-gas groups, and the collisions and mergers
of such objects as touched upon in this paper.

In the longer term, we plan to simulate larger more complex systems
(dynamics in turbulent gas or fractal distributions) both as simple
numerical experiments and to compare with observations.  
We plan to add simple stellar heating and gas cooling prescriptions 
in order to advance on the simple adiabatic EOS we have employed here.
For feedback and gas dispersal simulations, we will add some 
simple feedback formulations, like gas-heating from 
supernovae, as well as mechanical winds and UV radiation using
the HEALPix-based algorithm already implemented in SEREN \citep{Bisbas2009}.  
Simple accretion models can be added to allow stars to accrete from the gas.  

One important algorithmic addition we are currently implementing in the code is a more sophisticated N-body integrator that will allow efficient evaluation of computationally expensive sub-sytems such as tight binaries and 3- or 4-body encounters.  We are implementing an adaptive nearest-neighbour tree, similar to that used in the {\small STARLAB} N-body code suite \citep{starlab2001} {and {\small MYRIAD} \citep{MYRIAD}}, to decompose the stars into sub-systems and then use a higher-order integrator, such as the 6th or 8th order Hermite integrators, to more accurately integrate the sub-system.  This addition will allow us to model clusters containing primordial binaries and higher-order multiple systems, or clusters that form hard binaries.
Details and tests of any additional physics and optimisations will be 
explained in subsequent papers that introduce them.

\section{Conclusions} \label{S:CONCLUSIONS}  
  
We have presented a new hybrid SPH/$N$-body method within the SPH code
SEREN \citep{Hubber2011}.  Using conservative SPH and 4th-order
$N$-body integrators, this scheme conserves energy extremely well with an
adiabatic equation of state.  We have presented a number of tests of
the code showing that it works as expected in a number of simple
situations.  We will use this code in future to explore stellar
dynamics in a live gas background to investigate problems involving
star cluster formation and evolution.

\section*{Acknowledgements}  
DAH is funded by a Leverhulme Trust Research Project Grant (F/00 118/BJ) and an STFC post-doc, and was provided with a visitors grant through FONDECYT grant 1095092. RJA is supported by a research fellowship from the Alexander von Humboldt Foundation.  RS acknowledges support from FONDECYT grant number 3120135.  We thank the referee, Nickolas Moeckel, for some important comments and suggestions that helped to improve aspects of this paper.  We also thank Dr Daniel Price for making the SPLASH \citep{splash2007} code available, from which some of the figures in this paper were prepared.


\appendix

\section{Derivation of conservative SPH equations with stars} \label{A:SPHDERIVATION}  
  
Following \citet{PM2007}, we derive the SPH equations of motion, for both the gas and star particles, using Lagrangian mechanics.  For a set of $N_g$ gas particles with labels $b = 1,2, ... ,N_g$ and $N_s$ star particles with labels $i = 1,2, ... ,N_s$, then the Lagrangian becomes   
\begin{eqnarray} \label{EQN:LAGRANGIAN}  
{\cal L} &=& \frac{1}{2}\sum \limits_{b=1}^{N_g} {m\ssb \, v\ssb^2 }  + \frac{1}{2}\sum \limits_{i=1}^{N_s} {m\ssi\,v\ssi^2 } + {\cal L}_{_{\rm GRAV}}  
\end{eqnarray}  
where ${\cal L}_{_{\rm GRAV}}$ is the gravitational contribution to the Lagrangian given by   
\begin{eqnarray} \label{EQN:GRAVLAGRANGIAN}  
{\cal L}_{_{\rm GRAV}} &=&   
- \frac{G}{2} \sum \limits_{b=1}^{N_g} \, \sum \limits_{c=1}^{N_g}\, { m\ssb m\ssc \phi\ssbc(\overline{h}\ssbc) } -   
\frac{G}{2} \sum \limits_{i=1}^{N_s} \, \sum \limits_{j=1}^{N_s}\, { m\ssi m\ssj \phi\ssij(\overline{h}\ssij) }   
- G \sum \limits_{b=1}^{N_g} \, \sum \limits_{i=1}^{N_s}\, { m\ssb m\ssi \phi\ssbi(\overline{h}\ssbi) }   
\end{eqnarray}  
where $\overline{h}\ssbc \equiv \frac{1}{2}\left( h\ssb + h\ssc \right)$ is the mean smoothing length of particles $b$ and $c$.  The equations of motion can be obtained by inserting the Lagrangian into the Euler-Lagrange equations,   
\begin{eqnarray} \label{EQN:EULERLAGRANGE}  
\frac{d}{dt} \left( \frac{\partial {\cal L}}{\partial {\bf v}\ssa} \right)   
- \frac{\partial {\cal L}}{\partial {\bf r}\ssa} &=& 0  
\end{eqnarray}  
The Lagrangian is symmetric in terms of interaction terms between the gas and star particles.  The only difference lies in the method of calculating the smoothing length which leads to different forms of the equation of motion for both stars and gas.

\subsection{Gas particles}  
First, we derive the equation of motion for a general gas particle labelled $a$ by taking the derivative with respect to its position, ${\bf r}\ssa$, i.e.  
\begin{eqnarray}  
\frac{\partial {\cal L}_{_{\rm GRAV}}}{\partial {\bf r}\ssa} &=&  
-\frac{G}{2} \sum \limits_{b=1}^{N_g}\,\sum \limits_{c=1}^{N_g} {  
m\ssb\,m\ssc\,\left[ \phi'\ssbc(\meanhbc)\,\hat{\bf r}\ssbc\,(\delta\ssba - \delta\ssca)   
+ \frac{1}{2}\,\frac{\partial \phi\ssbc}{\partial \meanhbc}\,  
\left( \frac{\partial h\ssb}{\partial \rho\ssb}\frac{\partial \rho\ssb}{\partial {\bf r}\ssa} + \frac{\partial h\ssc}{\partial \rho\ssc}\frac{\partial \rho\ssc}{\partial {\bf r}\ssa} \right)  
\right] } \nonumber \\  
&& - G \sum \limits_{b=1}^{N_g}\,\sum \limits_{i=1}^{N_s} {  
m\ssb\,m\ssi\,\left[ \phi'\ssbi(\meanhbi)\,\hat{\bf r}\ssbi\,\delta\ssba  
+ \frac{1}{2}\,\frac{\partial \phi\ssbi}{\partial \meanhbi}\,  
\frac{\partial h\ssb}{\partial \rho\ssb}\frac{\partial \rho\ssb}{\partial {\bf r}\ssa} \right] }\,.  
\end{eqnarray}  
We note there is no contribution from the star-only term in the Lagrangian since there is no dependence on the position of any gas particles, i.e. ${\bf r}\ssa$.  Substituting the expression for $\partial \rho / \partial {\bf r}$, i.e.   
\begin{eqnarray} \label{EQN:GRADRHO}  
\frac{\partial \rho\ssb}{\partial {\bf r}\ssa} &=&  
\frac{1}{\Omega\ssb} \sum \limits_{d=1}^{N_g}   
{m\ssd \frac{\partial W\ssbd(h\ssb)}{\partial {\bf r}\ssa}\,  
(\delta\ssba - \delta\ssda) }\,,  
\end{eqnarray}  
where $\Omega\ssb$ is given by Equation \ref{EQN:OMEGA}, we obtain   
\begin{eqnarray}  
\frac{\partial {\cal L}_{_{\rm GRAV}}}{\partial {\bf r}\ssa} &=&  
-\frac{G}{2} \sum \limits_{b=1}^{N_g}\,\sum \limits_{c=1}^{N_g}   
m\ssb\,m\ssc\, \phi'\ssbc(\meanhbc)\,\hat{\bf r}\ssbc\,(\delta\ssba - \delta\ssca)   
- G \sum \limits_{b=1}^{N_g}\,\sum \limits_{i=1}^{N_s} m\ssb\,m\ssi\,  
\phi'\ssbs(\meanhbs)\,\hat{\bf r}\ssbs\,\delta\ssba \nonumber \\ && - \frac{G}{4}   
\sum \limits_{b=1}^{N_g}\,\sum \limits_{c=1}^{N_g}\,\sum \limits_{d=1}^{N_g}   
m\ssb\,m\ssc\,m\ssd \, \frac{\partial \phi\ssbc}{\partial \meanhbc}\,  
\left( \frac{1}{\Omega\ssb}\frac{\partial h\ssb}{\partial \rho\ssb}\frac{\partial W\ssbd(h\ssb)}{\partial {\bf r}\ssa}(\delta\ssba - \delta\ssda) + \frac{1}{\Omega\ssc}\frac{\partial h\ssc}{\partial \rho\ssc}\frac{\partial W\sscd(h\ssc)}{\partial {\bf r}\ssa}(\delta\ssca - \delta\ssda) \right)  \nonumber \\  
&& - \frac{G}{2}   
\sum \limits_{b=1}^{N_g}\,\sum \limits_{i=1}^{N_s}\,\sum \limits_{d=1}^{N_g}   
m\ssb\,m\ssi\,m\ssd \, \frac{\partial \phi\ssbs}{\partial \meanhbs}\,  
\frac{1}{\Omega\ssb}\frac{\partial h\ssb}{\partial \rho\ssb}\frac{\partial W\ssbd(h\ssb)}{\partial {\bf r}\ssa}(\delta\ssba - \delta\ssda)  
\end{eqnarray}  
Expanding out the Kronecker delta functions and simplifying,   
\begin{eqnarray}  
\frac{\partial {\cal L}_{_{\rm GRAV}}}{\partial {\bf r}\ssa} &=&  
-\frac{G}{2} \sum \limits_{c=1}^{N_g} m\ssa\,m\ssc\, \phi'\ssac(\meanhac)\,\hat{\bf r}\ssac + \frac{G}{2} \sum \limits_{b=1}^{N_g} m\ssb\,m\ssa\, \phi'\ssba(\meanhba)\,\hat{\bf r}\ssba  
- G \sum \limits_{i=1}^{N_s} m\ssa\,m\ssi\,  
\phi'\ssai(\meanhas)\,\hat{\bf r}\ssai   
\nonumber \\ && - \frac{G}{4}   
\sum \limits_{c=1}^{N_g}\,\sum \limits_{d=1}^{N_g}   
m\ssa\,m\ssc\,m\ssd \, \frac{\partial \phi\ssac}{\partial \meanhac}\,  
\frac{1}{\Omega\ssa}\frac{\partial h\ssa}{\partial \rho\ssa}\frac{\partial W\ssad(h\ssa)}{\partial {\bf r}\ssa}  
+ \frac{G}{4}   
\sum \limits_{b=1}^{N_g}\,\sum \limits_{c=1}^{N_g}   
m\ssb\,m\ssc\,m\ssa \, \frac{\partial \phi\ssbc}{\partial \meanhbc}\,  
\frac{1}{\Omega\ssb}\frac{\partial h\ssb}{\partial \rho\ssb}\frac{\partial W\ssba(h\ssb)}{\partial {\bf r}\ssa}  
 \nonumber \\  
&& - \frac{G}{4}   
\sum \limits_{b=1}^{N_g}\,\sum \limits_{d=1}^{N_g}   
m\ssb\,m\ssa\,m\ssd \, \frac{\partial \phi\ssba}{\partial \meanhba}\,  
\frac{1}{\Omega\ssa}\frac{\partial h\ssa}{\partial \rho\ssa}\frac{\partial W\ssad(h\ssa)}{\partial {\bf r}\ssa} + \frac{G}{4}   
\sum \limits_{b=1}^{N_g}\,\sum \limits_{c=1}^{N_g}  
m\ssb\,m\ssc\,m\ssa \, \frac{\partial \phi\ssbc}{\partial \meanhbc}\,  
\frac{1}{\Omega\ssc}\frac{\partial h\ssc}{\partial \rho\ssc}\frac{\partial W\ssca(h\ssc)}{\partial {\bf r}\ssa}  
  \nonumber \\  
&& - \frac{G}{2} \sum \limits_{i=1}^{N_s}\,\sum \limits_{d=1}^{N_g}   
m\ssa\,m\ssi\,m\ssd \, \frac{\partial \phi\ssai}{\partial \meanhai}\,  
\frac{1}{\Omega\ssa}\frac{\partial h\ssa}{\partial \rho\ssa}\frac{\partial W\ssad(h\ssa)}{\partial {\bf r}\ssa}   
+ \frac{G}{2} \sum \limits_{b=1}^{N_g}\,\sum \limits_{i=1}^{N_s}   
m\ssb\,m\ssi\,m\ssa \, \frac{\partial \phi\ssbi}{\partial \meanhbi}\,  
\frac{1}{\Omega\ssb}\frac{\partial h\ssb}{\partial \rho\ssb}\frac{\partial W\ssba(h\ssb)}{\partial {\bf r}\ssa} \nonumber \\  
&=&  
- G \sum \limits_{b=1}^{N_g} m\ssa\,m\ssb\, \phi'\ssab(\meanhab)\,\hat{\bf r}\ssab   
- G \sum \limits_{i=1}^{N_s} m\ssa\,m\ssi\,  
\phi'\ssas(\meanhai)\,\hat{\bf r}\ssai   
\nonumber \\ && - \frac{G}{2} \sum \limits_{b=1}^{N_g}   
m\ssa\,m\ssb \,\left[ \frac{\bar{\zeta}\ssa}{\Omega\ssa}\frac{\partial W\ssab(h\ssa)}{\partial {\bf r}\ssa}  + \frac{\bar{\zeta}\ssb}{\Omega\ssb}\frac{\partial W\ssab(h\ssb)}{\partial {\bf r}\ssa} \right]  
 - \frac{G}{2} \sum \limits_{b=1}^{N_g} m\ssa\,m\ssb \left[ \,\frac{\bar{\chi}\ssa}{\Omega\ssa} \frac{\partial W\ssab(h\ssa)}{\partial {\bf r}\ssa}   
+ \frac{\bar{\chi}\ssb}{\Omega\ssb}\frac{\partial W\ssab(h\ssb)}{\partial {\bf r}\ssa} \right]  
\end{eqnarray}  
where $\bar{\zeta}\ssa$ \citep[cf.][]{PM2007} and $\bar{\chi}\ssa$ are defined by   
\begin{eqnarray}  
\bar{\zeta}\ssa = \frac{\partial h\ssa}{\partial \rho\ssa}   
\sum \limits_{b=1}^{N} m\ssb \frac{\partial \phi\ssab}{\partial \meanhab}\,,  
\label{EQN:GRADHZETA2}  
\end{eqnarray}  
\begin{eqnarray}  
\bar{\chi}\ssa = \frac{\partial h\ssa}{\partial \rho\ssa}   
\sum \limits_{i=1}^{N} m\ssi \frac{\partial \phi\ssai}{\partial \meanhai}\,.  
\label{EQN:GRADHCHI2}  
\end{eqnarray}  
Substituting into the Euler-Lagrange Equation (Equation \ref{EQN:EULERLAGRANGE}), we obtain the equation of motion for SPH gas particles,  
\begin{eqnarray} \label{EQN:SPHEOM2}  
{\bf a}\ssa &=&   
- G \sum \limits_{b=1}^{N_g} m\ssb\, \phi'\ssab(\meanhab)\,\hat{\bf r}\ssab   
- G \sum \limits_{i=1}^{N_s} \,m\ssi\,  
\phi'\ssas(\meanhas)\,\hat{\bf r}\ssas   
- \frac{G}{2} \sum \limits_{b=1}^{N_g}   
m\ssb \,\left[ \frac{(\bar{\zeta}\ssa + \bar{\chi}\ssa)}{\Omega\ssa}\frac{\partial W\ssab(h\ssa)}{\partial {\bf r}\ssa}  + \frac{(\bar{\zeta}\ssb + \bar{\chi}\ssb)}{\Omega\ssb}\frac{\partial W\ssab(h\ssb)}{\partial {\bf r}\ssa} \right]\,.  
\end{eqnarray}

\subsection{Star particles}  
  
Similarly for star particles, we derive the equation of motion for a general star particle labelled $a$ by taking the derivative of the gravitational component of the Lagrangian with respect to its position, ${\bf r}\ssa$, i.e.  
\begin{eqnarray}  
\frac{\partial {\cal L}_{_{\rm GRAV}}}{\partial {\bf r}\ssa} &=&  
-\frac{G}{2} \sum \limits_{i=1}^{N_s}\,\sum \limits_{j=1}^{N_s} {  
m\sss\,m\ssj\,\left[ \phi'\ssij(\meanhij)\,\hat{\bf r}\ssij\,(\delta\ssia - \delta\ssja) \right] }   
 - G \sum \limits_{b=1}^{N_g}\,\sum \limits_{i=1}^{N_s} {  
m\ssb\,m\ssi\,\left[ \phi'\ssbi(\meanhbi)\,\hat{\bf r}\ssbi\,(-\delta\ssia) \right] } \nonumber \\  
&=& -\frac{G}{2} \sum \limits_{j=1}^{N_s} {  
m\ssa\,m\ssj\,\phi'\ssat(\meanhaj)\,\hat{\bf r}\ssaj} \,  
+ \frac{G}{2} \sum \limits_{i=1}^{N_s}\,{  
m\ssi\,m\ssa\, \phi'\sssa(\meanhia)\,\hat{\bf r}\ssia }\,  
 + G \sum \limits_{b=1}^{N_g}\,{  
m\ssb\,m\ssa\,\phi'\ssba(\meanhba)\,\hat{\bf r}\ssba }\,  
\nonumber \\  
&=& -G \sum \limits_{i=1}^{N_s} {  
m\ssa\,m\ssi\,\phi'\ssai(\meanhai)\,\hat{\bf r}\ssai} \,  
 - G \sum \limits_{b=1}^{N_g}\,{  
m\ssa\,m\ssb\,\phi'\ssab(\meanhab)\,\hat{\bf r}\ssab }\,  
\end{eqnarray}  
Due to the stars having constant smoothing length, we obtain somewhat simpler equations than for the case of gas particles.  Substituting into the Euler-Lagrange equations and renaming some summations for clarity, we obtain the following expression for the acceleration of star $s$,  
\begin{eqnarray}  
{\bf a}\sss  
&=& -G \sum \limits_{i=1}^{N_s} {  
m\ssi\,\phi'\sssi(\meanhsi)\,\hat{\bf r}\sssi} \,  
 - G \sum \limits_{b=1}^{N_g}\,{  
m\ssb\,\phi'\sssb(\meanhsb)\,\hat{\bf r}\sssb }\,  
\label{EQN:STARGRAV2}  
\end{eqnarray}

\end{document}